\documentclass[preprint,times,12pt]{elsarticle}

\graphicspath{ {./images/} }
\usepackage{amssymb}
\usepackage{amsmath}
\usepackage{caption}
\usepackage[parfill]{parskip}
\usepackage{subcaption}
\usepackage{url}
\usepackage{multicol}
\usepackage[ruled, lined, linesnumbered, commentsnumbered, longend]{algorithm2e}
\usepackage{nomencl}

\newcommand{\pushright}[1]{\ifmeasuring@#1\else\omit\hfill$\displaystyle#1$\fi\ignorespaces}

\journal{Applied Energy}

\begin{document}

\begin{frontmatter}

\title{Robust Market Potential Assessment: Designing optimal policies for low-carbon technology adoption in an increasingly uncertain world}

\affiliation[imp]{organization={Centre for Process Systems Engineering},
            addressline={Imperial College London},
            city={London},
            postcode={SW7 2BB}, 
            state={London},
            country={United Kingdom}}
            
\affiliation[cep]{organization={Centre for Environmental Policy},
            addressline={Imperial College London},
            city={London},
            postcode={SW7 1NE}, 
            state={London},
            country={United Kingdom}}

\author[imp]{Tom Savage}
\author[imp]{Antonio del Rio Chanona}
\author[cep]{Gbemi Oluleye}\ead{o.oluleye@imperial.ac.uk}

\newpage
\begin{abstract}
Increasing the adoption of alternative technologies is vital to ensure a successful transition to net-zero emissions in the manufacturing sector.
Yet there is no model to analyse technology adoption and the impact of policy interventions in generating sufficient demand to reduce cost.  
Such a model is vital for assessing policy-instruments for the implementation of future energy scenarios. 
The design of successful policies for technology uptake becomes increasingly difficult when associated market forces/factors are uncertain, such as energy prices or technology efficiencies. 
In this paper we formulate a novel robust market potential assessment problem under uncertainty, resulting in policies that are immune to uncertain factors. 
We demonstrate two case studies: the potential use of carbon capture and storage for iron and steel production across the EU, and the transition to hydrogen from natural gas in steam boilers across the chemicals industry in the UK. 
Each robust optimisation problem is solved using an iterative cutting planes algorithm which enables existing models to be solved under uncertainty. 
By taking advantage of parallelisation we are able to solve the nonlinear robust market assessment problem for technology adoption in times within the same order of magnitude as the nominal problem. 
Policy makers often wish to trade-off certainty with effectiveness of a solution. 
Therefore, we apply an approximation to chance constraints, varying the amount of uncertainty to locate less certain but more effective solutions. 
Our results demonstrate the possibility of locating robust policies for the implementation of low-carbon technologies, as well as providing direct insights for policy-makers into the decrease in policy effectiveness that results from increasing robustness. 
The approach we present is extensible to a large number of policy design and alternative technology adoption problems.
\end{abstract}


\begin{highlights}
\item A novel alternative technology adoption model for the manufacturing sector.
\item Formulation of market potential assessment problem under uncertainty.
\item Application of state-of-the-art cutting-set robust optimisation method.
\item Insights into robust policy interventions to support low carbon technologies.
\end{highlights}


\end{frontmatter}

\newpage

\makenomenclature

\printnomenclature
\newpage

\section{Introduction}

\subsection{Modelling Alternative Technology and Fuel Adoption and the Impact of Policy Interventions}

Difficult to abate sectors such as manufacturing contribute to 20\% of global carbon emissions \citep{worldbank}.
Achieving the goals of the Paris agreement will require decarbonising the industrial sector through the uptake of alternative technologies such as carbon capture and storage, fuel cells, and heat pumps, as well as the uptake of alternative sources of energy such as hydrogen, biofuels, and green electricity.
Within the manufacturing sector, the production of iron and steel, chemicals, and minerals account for 44\% of total industry-based carbon emissions. 
A major barrier to the adoption of alternative technologies and fuels in the industrial sector are the high associated investment costs \citep{European, Napp2014}.
Alternative technologies and fuels may experience a technology ``valley of death'' in which technology and market risks remain high while investments are low. 
This implies that interventions in the form of policy support are required to accelerate the uptake of these low-carbon alternatives to enable industry to comply with the Paris Agreement. 
Several studies have agreed on the need for policy support for industry \citep{hman2016,Binetti2023,Vieira2021} and determined the policy typologies for the manufacturing sector \citep{Rissman2020}; yet determining the optimal mix of policies, and their impact on reducing cost and generating demand to increase adoption requires further investigation.

Further down the value chain, the modelling of consumer technology adoption, despite being inherently anthropological, stochastic, and therefore challenging, is more common. 
In understanding energy-related consumer decision making, previous authors have integrated behavioural and social phenomena, trust-based information networks, and social norms.
To model consumer technology adoption, standard models such as Bass diffusion models, dynamic discrete choice models, and conjoint analysis have been proposed alongside more recent agent-based models. 
Dynamic discrete choice models are considered to be the most sophisticated for analysing consumer choice for alternative technology adoption as they allow for the study of individual decision-making processes on system outcomes \citep{Aguirregabiria2010}.
However, these predominant methods for modelling consumer technology adoption are not relevant for organisational decision making, particularly for decision making within the manufacturing sector.
As organisations are expected to act more rationally, and therefore generally require more certainty, or equivalently an understanding of uncertainty, there is a greater need to model the adoption of new technologies.

Organisations rely heavily on actors who have irrational expectations about technology trajectory compared to rational expectations for technology manufacturing. 
The complexity of manufacturing means that technologies need to be integrated systematically and uncertainties need to be accounted for \citep{Sunny2022,Cobo2023}. 
Furthermore, conventional models do not demonstrate how increased demand can reduce cost by exploiting learning-by-doing, economies of scale, and learning-by-innovation especially for technologies and fuels with low readiness levels. 
Although alternative technology and fuel adoption has attracted the attention of academics and politicians in recent times, there is no systematic and inclusive model of clean technology adoption for organisations within the manufacturing sector \citep{Dincbas2021}. 
Such a model would require a more complex formulation that accounts for uncertainty in both the technology and its end-use. 
This paper seeks to develop a novel robust market potential assessment (MPA) model for modelling the adoption of alternative technologies and fuels in the manufacturing sector. 
It is based on a hypothesis that the size of demand i.e., a market can be exploited to drive down cost reduction, and adoption can be accelerated with the right policies made under uncertainty. 
The robust MPA is the first to produce policies that are guaranteed to demonstrate technology cost reduction from increased demand, regardless of uncertain factors. 
Policy interventions for technology adoption can be broadly divided into incentives and taxes; both of which are accounted for in this paper \citep{hman2016,RichardsonBarlow2022}.

\subsection{Modelling Uncertainties in Technology and Fuel Adoption}

Technology investment cost distributions from numerous engineering studies of nuclear, biomass, solar and wind technologies demonstrate that technology change and learning effects are highly uncertain \citep{Lambert2019}. 
The presence of uncertainty makes the availability, adoption, and diffusion of alternative technologies slower and more discontinuous than would be if these decisions were made using nominal models.
In practice, stakeholders passively seek to realise unknown uncertainties by waiting for them to instantiate, resulting in slower adoption. 
Traditional deterministic optimisation and social planning models have been criticised for being optimistic in their handling of uncertainties and technologically too pessimistic by neglecting reduction in technology cost over time from increased demand and spillover effects \citep{Ma2009}. 
These stated downsides have contributed to an increasing number of nonlinear stochastic and robust problem formulations to account for these known uncertainties. 

Uncertainties in price fluctuations have been addressed using a discrete stochastic autoregressive moving average process \citep{Madlener2005}. 
In \citet{Karan2016} a stochastic approach was used to design the adoption of a sustainable food-energy-water systems. 
\citet{Zeng2020} considered uncertainties in the adoption of end-of-pipe abatement technologies and the effect of emission taxes and standards using sensitivity analysis. 
The impact of uncertainties when modelling and optimising for policy interventions to support the adoption of alternative technologies in decarbonising the manufacturing industry has yet to be assessed. 

Given the increasingly uncertain nature of energy systems, as a result of geopolitical factors, this uncertainty is becoming increasingly important to account for. 
This uncertainty that occurs in technology adoption and energy policy models can occur as a result of the nature of the system at different scales\citep{Liu2022,Liu20222,Li2022}.
For example, micro-scale stochasticity in the supply and demand of energy throughout an energy grid \citep{Rahim2022}, or on a macro-level the uncertainty in the effectiveness of future energy technologies \citep{Ahmed2014}.
Within energy policy, it is important to account for all aspects of uncertainty, and communicate their impacts to policy makers \citep{Workman2020,Bushell2017}.
Recent emphasis has been placed on communicating the impacts of these macro, `deep' uncertainties \citep{Workman2020}.
Throughout the UK committee for climate change's independent assessment on the UK's heat and building strategy, policies are assigned a feasibility from credible to significant risk \citep{CCC2018}. 
In an ideal world, policy makers should enact policies that are feasible whilst still accounting for all aspects of uncertainty, ever increasing. 
This naturally leads to the concept of providing robust solutions to policy-focused energy systems models under uncertainty. 

Stochastic optimisation provides an alternative solution to optimisation under uncertainty, albeit entailing a different philosophy.
Instead of optimising for the worst-case scenario within a set, solutions are found with respect to the probability distribution of uncertain parameters, satisfied a percentage of the time.
Inherent benefits include the ability to include distributional information regarding uncertain parameters, when this information is available \citep{Abunima2022}. 
However, difficulties arise in the generation of realistic scenarios.
Likewise, defining the distribution of uncertain parameters can rely on assumptions and often historical data.
Distributionally robust optimisation provides a common ground between stochastic and robust optimisation \citep{DRO,DRO2}. 
Uncertain optimisation problems are solved in expectation for worst-case distributions of uncertain parameters. 
However, distributionally robust optimisation has been known to be heavily reliant on historical datasets to produce accurate distributional ambiguity sets \citep{DRO}. 
Within energy policy problems, data exists only for some parameters, whereas parameters such as technology learning rates or the efficiency of future technologies are yet to progress beyond bounded estimates.  
For energy system problems with known uncertainties such as the stochastic demand of electricity, distributional information may be available, in which case distributionally robust or stochastic optimisation is a promising approach.

Within robust optimisation there are differing philosophies in how to approach gaining a feasible solution under uncertainty, particularly when presented with a nonlinear or non-convex model as energy policy models often are.
As opposed to directly finding robust optimal solutions to nonlinear or non-convex robust optimisation problems, the model itself may be simplified to enable reformulation \citep{Yliruka2022,Moret2020}.
As such, the suite of tools available for linear robust optimisation can be deployed \citep{BenTal2014}. 
However, in this study instead of making simplifying assumptions, we choose to maintain the original market potential assessment problem formulation. 
This enables us to consider important nonlinear market effects such as the learning rates of low-carbon technologies. 

\textbf{In this paper} we identify robust policies to problems concerning the transition to low-carbon technologies.
Our approach can accelerate uptake of alternative technologies and fuel in the industrial sector.
The work presented in this paper bridges the gap between research, development and implementation by presenting a novel robust market potential assessment model to study and analyse technology adoption in the manufacturing sector considering uncertainty in all the model’s inputs.
The robust market potential assessment model is the first to exploit how market size together with policy interventions can create sufficient demand to trigger and sustain technology cost reduction to accelerate uptake. 

\subsection{Related Work}
Modelling for alternative fuel and technology adoption is an important part of enabling the transition to net zero. 
However, the most common modelling techniques in literature focus on consumer uptake of technology. Examples of common techniques are agent-based models and discrete choice modelling \citep{Maybury2022}. 
Agent-based models are simulation-based, and provide a quantitative depiction of technology adoption in the context of exogenous scenarios. 
In contrast, discrete choice models are equation-based accounting models. 
These models can represent the inventory of technologies adopted under the influence of existing policy interventions but have not been used to design new policies to accelerate technology adoption under uncertainty. 

Agent-based models have been applied to the diffusion of electric or plug-in electric vehicles \citep{Silvia2016}, adoption of residential solar PV systems with battery systems \citep{Alyousef2016,Rai2015}, and adoption of heat pumps \citep{Meles2022,Savage2022heatpump}. 
The 23 agent-based models reviewed by \citet{Hesselink2019} focused on energy efficiency technology adoption by households. 
The focus of these studies is largely on consumer adoption, as opposed to policy design and quantifying the impact of policies.
Furthermore, these models do not integrate technology learning from doing, hence the impact of learning-by-doing on cost reduction due to policies is seldom estimated. 
Even though these models integrate empirically driven agent interaction and theoretically driven behavioural models, they at best replicate past consumer markets.
The findings of agent-based modelling research is still useful to guide policy makers and companies in devising monetary and non-monetary interventions aimed at increasing future uptake of alternative technologies.
However, models that can design policies under uncertainty and quantify their impact in generating sufficient market demand for a technology to trigger cost reduction would be more useful, especially for the manufacturing sector. 
Organisational decision making is more structured and subject to finding an economic business case, yet there is no model suitable for technology adoption within this space.

A unique feature of adoption of technology in manufacturing is the heterogeneous nature of each organisation/industrial site. 
Even though agent-based models account for heterogeneous agents with bounded rationality, the ability to reduce cost from increased demand from the heterogeneous agent using the right incentives is not often determined. 
Furthermore, agent-based models have not been used to determine the optimal adoption curve given uncertainties in both exogenous and endogenous variables. 
Therefore, the application of mathematical modelling is important as it allows one to formulate and solve these adoption problems by drawing from existing knowledge.
Of the 55 models reviewed by \citet{Maybury2022}, none exploited the market size together with interventions to reduce technology cost. 
Therefore, there is need for an equation-based optimisation model capable of modelling high-level heterogeneous technology adoption and complexity of decision making by organisations. 

Another class of model for adoption are system dynamics, evolutionary game frameworks, and market-based models. 
\citet{Hsu2012} applied system dynamics to assess the impact of policies in solar PV installations. 
The model allows policy makers to undertake cost/benefit analyses for different combinations of policies. 
Neither technology learning nor the potential of market size were integrated, and no uncertainty was considered. 
\citet{Simonsen2022} applied a scenario-based tool to compare different policy approaches in the context of household energy use in Norway. 
The first model to address technology adoption was developed by \citet{Zhao2019} for carbon capture and storage (CCS) adoption at the micro-level using a government enterprise evolutionary game framework. 
However, the study considered a relatively small market, did not consider the impact of policies, and cannot be used to determine a combination of policies to support adoption. 

There is need to develop models for adoption from a market perspective instead of a consumer perspective as this is more relevant for manufacturers \citep{Khanam2021}. 
An important element of market-derived technology adoption is technology learning; learning-by-innovation, learning-by-doing and economies of scale help determine when adoption of a technology causes market saturation i.e., can the size of the market be exploited together with policies to reduce cost of technology? 
Answering these questions would require integrating technology learning into a policy framework. 
Technology learning, as it is broadly concerned with the future, is inherently uncertain.
As such, any market-based model needs to be developed to handle uncertainties in both technology learning, and other aspects of the system.  
\citet{Liu2022sep} applied a component-based two factor technology learning curve approach to predict the future cost as part of a new real option investment model development. 
The impact of increased demand due to effective policies was not considered. 
The modelling in \citet{Liu2022sep} integrated least square Monte Carlo simulations to handle uncertainty. 
\citet{Guo2021} applied combined Monte Carlo and random search methods to solve a stochastic dynamic optimisation problem, demonstrating the ability to handle multiple complexities in the technology adoption process. 
\citet{Stavrakas2019} presented an agent-based technology adoption model under uncertainty. 
However, this agent-based model quantified only uncertainties related to agents' preferences using sensitivity and scenario analysis. 

Uncertainty is ubiquitous in alternative technology and fuel adoption, particularly when considering future technologies with unrealised uncertainties such as carbon capture and storage and hydrogen. 
Stochastic programming and robust optimisation are among the major approaches for addressing this uncertainty. 
\citet{Xun2022} compared deterministic and robust optimisation for the design and optimisation of fuel cell hybrid electric trucks. The authors assign a low probability of violation for stochastic constraints, approximating the true robust solution. 
Robust optimisation has not been applied in the context of technology adoption, despite its attractive paradigm in optimising over the worst-case scenarios. 
Robust optimisation has been used within the context of process systems engineering to address many open questions in alternative technology design, integration, and assessment but not adoption.
\citet{Riepin2022} investigates adaptive robust optimisation for gas infrastructure planning. 
The authors apply an iterative constraint and column generation algorithm to solve the problem.
Process systems engineering has advanced various technology related applications from design to supply chain, with the majority of the mathematical programming approaches including mixed integer non-linear programming, multi-objective optimisation and Monte Carlo based algorithms. 
Mathematical programming has been applied to investigate the impact of policy on alternative technology. 
For example, \citet{Fan2018} explored 45Q tax credit for CCS integration in existing coal-fired power plants, \citet{Quarton2021} explored carbon budgets and carbon taxation in achieving net-zero emissions, and \citet{Ratanakuakangwan2022} solved a multi-objective energy planning model to identify optimal energy policies. 
Designed policies have been shown to significantly influence overall costs and adoption. 
\citet{Chapman2020} developed a global energy economic optimisation model to determine the potential societal penetration of hydrogen integrated technology learning curves. 
However, the learning of a technology was only applied to generation technologies for hydrogen, but not for the demand of hydrogen. 
The sensitivity of these policies was investigated across different scenarios of carbon tax rate. 

No study to date has extended robust optimisation to the field of technology adoption. 
Additionally, none of these studies design and evaluate policies and the role they play in generating market demand that could trigger technology cost reduction in order to increase uptake.
Technology adoption can effectively reduce mitigation costs due to ``learning-by-doing'', due to cumulative deployment. 
Technology learning happens due to the accumulation of experience, knowledge, and utilisation.
Increased demand is driven by policy interventions, and with it comes the accumulation of utilisation due to the noncompetitive nature of alternative fuels and technology for decarbonising the manufacturing sector. 
By exploiting policy interventions in the manufacturing sector to stimulate demand for technology adoption, our approach can ease the difficulty in achieving global and regional long-term emission mitigation targets.
Accounting for the uncertainty of technological learning in a robust framework can result in policies certain to reduce technology cost and therefore increase technology uptake rate. 

\subsection{Contribution}

The contributions of this article are three-fold. 
We first present novel robust market potential assessment model to study and analyse technology adoption in the manufacturing sector considering uncertainty in all the model parameters, and taking into account important effects such as technology learning.
The robust market potential assessment model is the first to exploit how market size together with policy interventions can create sufficient demand to trigger and sustain technology cost reduction to accelerate uptake. 
Secondly, we apply a state-of-the-art iterative robust optimisation algorithm and subsequently investigate robust solutions to the model, which immunise a solution against all aspects of uncertainty. 
This approach enables us to maintain the original model formulation, including nonlinear effects such as the learning rate of a technology, whilst providing an approximate robust solution. 
We investigate the effect of worst-case uncertainty on the model, providing conclusions regarding the adoption of an alternative technology such as CCS, and transition from natural gas to hydrogen under different scenarios. 
Finally, by making assumptions as to the form of uncertainty we are able to reduce the conservatism of the robust solution at the expense of increased risk. 
The insights gained by reducing the conservatism of the problem provides direct insights for policy makers, including the probability a solution will be made infeasible under some assumptions. 
The approach we present results in an approximate solution to worst-case uncertainty for a given nonlinear model. 
We believe this approach is more beneficial than making simplifying assumptions to gain an exact solution to worst-case uncertainty, particularly for technology adoption related policy optimisation. 
Our approach is general enough to be extended to different technologies and fuels.

\section{Methodology}
In this section we present problem formulations concerning robust market potential assessment of carbon capture and storage for emissions abatement in the iron and steel industry, as well as hydrogen adoption for boilers across the chemicals sector.
We first outline our approach for solving these problems robustly, highlighting the specific benefits of this approach. 

\subsection{Robust Optimisation}
\begin{figure}[htb!]
    \centering
    \includegraphics[width=\textwidth]{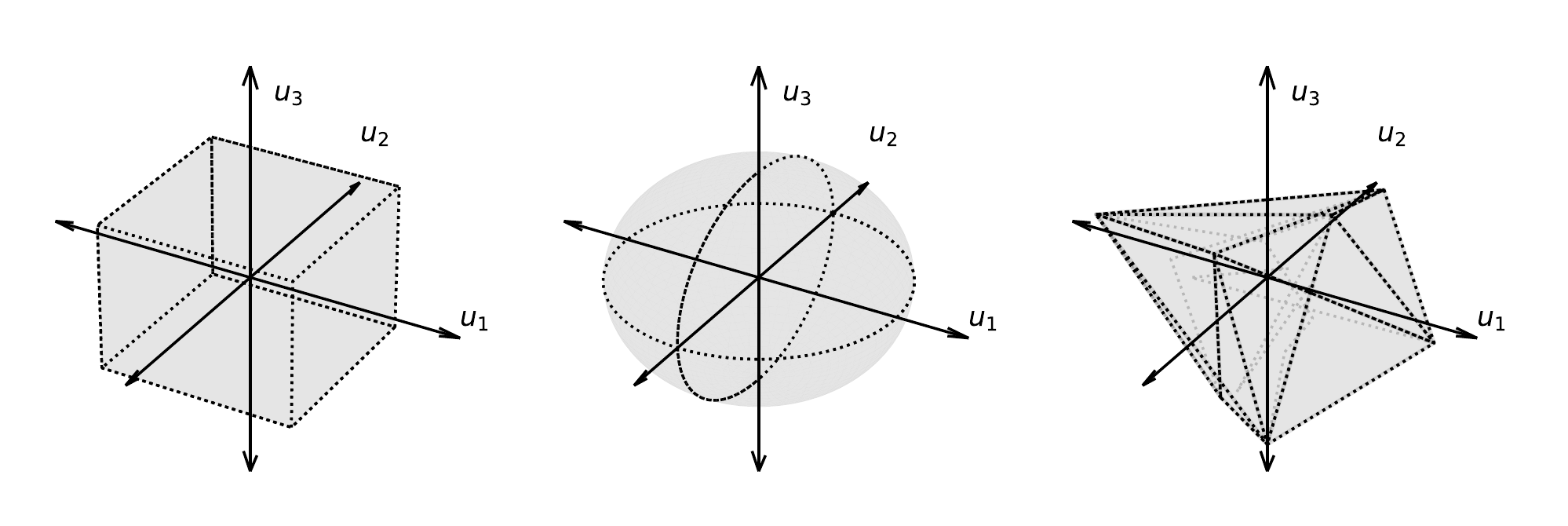}
    \caption{Three different uncertainty sets which represent the potential values that uncertain parameters can take. Left: a box which is the most conservative as all parameters may be simultaneously at their extreme values within a corner. Centre: an ellipse which is considered less conservative than the box. Right: a polyhedron which may be inferred from data.}
    \label{unc_set}
\end{figure}

The goal of robust optimisation is to make a decision that is feasible in the worst-case scenario. In other words, it aims to find solutions to optimisation problems such that constraints are satisfied for all potential values of uncertain parameters \citep{BEN09}.
Potential values of uncertain parameters are defined by an uncertainty set, a continuous set which contains all realisable instances of uncertainty.
To account for uncertain parameters that appear in the objective function, the addition of an additional variable and constraint means the problem can be converted into its epigraph form, such that the objective becomes deterministic \citep{BEN09}.
The canonical robust optimisation problem is therefore as follows:

\begin{align}
\min_{\mathbf{x}\in\mathcal{X}\subseteq\mathbb{R}^n} &\quad f(\mathbf{x}) \label{rob_obj}\\
    \text{s.t.} & \quad g_i(\mathbf{x},\mathbf{u}) \leq 0  \quad \forall \mathbf{u} \in \mathcal{U}\subseteq \mathbb{R}^m \quad i=1,\dots,I\label{rob_con}
\end{align}

where $\mathbf{x}$ are $n$ decision variables bounded by $\mathcal{X}$, and $\mathbf{u}$ are $m$ uncertain parameters, bounded by the uncertainty set $\mathcal{U}$.
The difficulty in robust optimisation arises from the fact that uncertain constraints must be satisfied under an infinite number of realisations, in the case that $\mathcal{U}$ is continuous. 
Therefore, robust optimisation has strong links to the field of semi-infinite programming \citep{Djelassi2021}.
In the case that an uncertain constraint $g(x,u)$ is linear in $x$ and $u$, and the uncertainty set satisfies certain conditions, the semi-infinite constraint can be reformulated into a finite number of equivalent counterparts \citep{BenTal2014, BEN09}.
When $g(x,u)$ is convex in $x$ and concave in $u$, the semi-infinite constraint can be reformulated into a tractable equivalent through the use of the Fenchel dual \citep{BenTal2014}.
These reformulations are highly dependent on the uncertainty set. 
For example, in the case that $g_i(\mathbf{x},\mathbf{u})$ is linear, a simple `box' uncertainty set which specifies upper and lower bounds for each parameter results in 4 equivalent linear constraints and $m$ additional variables. 
When an ellipsoid-shaped uncertainty set is applied, the reformulation consists of a single second-order conic constraint. 
Figure \ref{unc_set} demonstrates different shaped uncertainty sets, each of which results in different reformulations.

In parallel to reformulation-based approaches, approaches that approximate the infinite number of constraints by a finite subset have been developed.
These methods result in the following tractable approximation to Eq. \ref{rob_obj} and \ref{rob_con}:
\begin{align}
    \min_{\mathbf{x}\in\mathcal{X}\subseteq\mathbb{R}^n} &\quad f(\mathbf{x}) \\
    \text{s.t.} & \quad g_i(\mathbf{x},\mathbf{u}) \leq 0  \quad  \forall \mathbf{u} \in \mathcal{U}_k \quad i=1,\dots,I
\end{align}
where $\mathcal{U}_k:=\{\mathbf{u}_1,\dots,\mathbf{u}_k\}$ is the finite approximation of uncertainty set $\mathcal{U}$.
A method for semi-infinite programming that relies on this concept was first published by Blankenship \& Faulk \citep{Blankenship1976}, and more recently has been characterised under cutting-set, outer-approximation, or pessimisation-based approaches \citep{Djelassi2021}.
The algorithm is described as follows.
$\mathcal{U}_k$ is first initialised with a single value of $\mathbf{u}$.
The problem is initially solved with nominal parameters (i.e. $\mathcal{U}_k = \{\mathbf{u}_0\}$) to yield an initial solution $\mathbf{x}^*$.
Subsequently, for all constraints, a pessimisation problem, 
\begin{align}
   \mathbf{u}^* = \arg\max_{\mathbf{u}\in\mathcal{U}} g(\mathbf{x}^*,\mathbf{u}),
\end{align}
is solved, to locate `worst-case' values of uncertain parameters with respect to a previously solution $\mathbf{x}^*$. 
Worst-case parameters $\mathbf{u}^*$ are then added to $\mathcal{U}_k$ and the process repeated.
In recent years this type of approach has been shown to perform competitively with reformulation \citep{Mutapcic2009}, particularly in the case that $\mathcal{U}$ is an ellipse. 
Regardless of competitiveness with reformulation, a key advantage of this approach is that it can be applied in situations when the problem cannot be reformulated.
Other advantages include: reduced dependency on a specific uncertainty set, the ability to express a problem using its original formulation, and the ability to take advantage of computational speed-ups including parallelisation, warm-starting.
Overall with these advantages, cutting-set/outer approximation methods have been shown to demonstrate a multiplicative increase in solution time over the deterministic problem \citep{Mutapcic2009}. 
When considering the life-cycle from problem formulation to robust solution, these approaches result in even greater time savings over a reformulation based approach, enabling fast prototyping and application to existing problems. 
Therefore, this approach is the one that has been adopted in this article for the solution of robust energy policy problems with nonlinear constraints.

\subsection{Problem statement}
The Robust MPA is a model for understanding technology adoption under uncertainty.
We present two problem formulations.

\subsubsection{Robust MPA for Alternative Technologies}

Alternative technologies are crucial for emissions abatements, and achieving carbon neutrality targets in the manufacturing sector. 
However, high capital, operating and maintenance costs as well as high uncertainties in energy prices hinder their development. 
Examples of alternative technologies are: carbon capture and storage, fuel cells, heat pumps, and electric boilers. 
In this problem formulation, we investigate the market effects on the uptake of an alternative technology.  
The policy instruments are designated as decision variables. 
The objective is to minimise the difference between the initial cost of the technology and the final cost determined due to the impact of increased demand from policy interventions, subject to the creation of a market for the technology. 
This implies a minimisation in the policy incentives required to stimulate this market.
The final cost is a function of the technology learning rate. 
Equations \ref{start} to \ref{end} describe the problem. 

\nomenclature{\(E_i\)}{$\text{CO}_2$ emissions associated with plant $i$ (t/yr)}
\nomenclature{\(\eta\)}{Technology efficiency (t/t)}
\nomenclature{\(t_i\)}{Carbon Tax (\$/t) for plant $i$}
\nomenclature{\(m_i\)}{Market share of alternate technology for plant $i$}
\nomenclature{\(C_0\)}{Initial cost of technology per unit emissions reduced (\$/t)}
\nomenclature{\(A_0\)}{Initial emissions reduction (t/yr)}
\nomenclature{\(L_r\)}{Learning rate of technology}
\nomenclature{\(M_U\)}{Market Share Limit}

\begin{align}\label{start}
    \min_{\mathbf{t},\mathbf{m}} \quad & C_0\left[1-\left(\frac{A_0+\eta\sum_{i=1}^Im_i\sum_{i=1}^IE_i}{A_0}\right)^{-LP}\right] \\ \label{con1}
    \text{s.t.} \quad & L_P = {\frac{\log(1-L_r)}{\log(2)}}\\\label{con}
    & E_i(C_0 -\eta t_i) \leq 0 \quad i=1,\dots,I\\
    & m_i -  \frac{E_i}{\sum^I_{i=1}E_i} \leq 0 \quad i=1,\dots,I\\
    & M_U-\left[1-\left(\frac{A_0+\eta\sum_{i=1}^Im_i\sum_{i=1}^IE_i}{A_0}\right)^{L_P}\right]\leq 0 .\label{end}
\end{align}

In Eqs. \ref{start} to \ref{end}, vectors $\mathbf{t}=\{t_1,\dots,t_I\}$ and $\mathbf{m}=\{m_1,...,m_I\}$ represent the carbon tax and market share associated to each carbon producing process respectively.
The set of processes is represented by $I$ with individual processes indexed by $i$.
The scalar $\eta$ is the efficiency of alternative technology. 
Likewise scalars $C_0$, $A_0$ and $L_r$ represent the initial cost of alternate technology, the initial emissions reduced by alternate technology, and the learning rate of the technology respectively. 
$M_U$ is a parameter representing the market share limit of the transition technology, dictated by existing policy. 

To investigate optimal solutions to the problem under uncertainty, we assume that all parameters are uncertain.  
These include the reported CO\textsubscript{2} emissions of each generating process $E_i$, the CO\textsubscript{2} mitigation rate provided for by the alternate technology $\eta$, and the learning rate $L_r$ of the technology amongst others.  
Therefore constraints appearing in Equations \ref{con} to \ref{end} must be satisfied for all values of these uncertain parameters to satisfy the robust problem.
Note the constraint in Equation \ref{con1} will be subsumed into the remaining inequality constraints.
The market share limit, $M_U$ is a hyper-parameter which is subject to independent investigation. 

The resulting robust market potential assessment problem is as follows: 
\begin{align}\label{start_rob}
    \min_{\textbf{t},\textbf{m},\tau} \quad & \tau \\
    \text{s.t.} \quad & C_0\left[1-\left(\frac{A_0+\eta\sum_{i=1}^Im_i\sum_{i=1}^IE_i}{A_0}\right)^{-LP}\right] - \tau \leq 0 \quad \forall u\in\mathcal{U} \\\ 
     & L_P = {\frac{\log(1-L_r)}{\log(2)}}\\
    & E_i(C_0 -\eta t_i) \leq 0 \quad \forall u\in\mathcal{U}  \quad i=1,\dots,I \\
    & m_i -  \frac{E_i}{\sum^I_{i=1}E_i} \leq 0 \quad \forall u\in\mathcal{U}  \quad i=1,\dots,I\\
    & M_U-\left[1-\left(\frac{A_0+\eta\sum_{i=1}^Im_i\sum_{i=1}^IE_i}{A_0}\right)^{L_P}\right]\leq 0 \quad \forall u\in\mathcal{U}
\end{align}

where $\tau$ is the epigraph variable introduced to maintain a deterministic objective, $u$ denotes the following set of uncertain parameters $\mathbf{u}:=\{A_0,\eta,L_r,C_0,\mathbf{E}\}$, each with respective upper and lower bounds which we denote as the set $\mathbf{b}$.
$\mathcal{U}$ is the uncertainty set defining the values these uncertain parameters can take.
For example, in the case that the uncertainty set is a box, $\mathcal{U}$ is defined as $\mathbf{b}_1\times\dots\mathbf{b}_m$.

\subsubsection{Robust MPA for an Alternative Fuel Adoption}

Replacing fossil fuels with alternatives like hydrogen, biogas, synthetic methane is vital to reduce emissions throughout the manufacturing sector. 
Alternative fuels are more expensive than the incumbent; hence the need for policy interventions is crucial. 
Demand for fuels from the manufacturing sector may be sufficient to reduce cost; and policies can be designed to increase demand. 
In this sector we formulate the robust market potential assessment for adoption of an alternative fuel in the process industry. 
The policy instruments are designated as decision variables. 
The objective is to minimise the difference between the incumbent cost of the alternative fuel, and the new cost as a result of increased demand from policy interventions. 
The new cost is determined using the learning rate. 
An alternative fuel can become cost competitive with the incumbent when taking into account market effects such as learning rate.
Equations \ref{start2} to \ref{end2} describe the problem. 
Similar to the previous case study, we make the assumption for the purposes of investigation that all parameters are inherently uncertain.
Uncertain parameters in this case include the energy consumption data reported for a given manufacturing site $W_j$, alongside the learning rate of the fuel $L_r$ and conversion technology efficiency $\eta_{F_2}$.
As before, constraints appearing in Equations \ref{start2} to \ref{end2} must be satisfied for all values of these uncertain parameters to satisfy the robust problem. 
We present the nominal formulation as follows:

\nomenclature{\(LHV_{F_1}\)}{Initial Fuel LHV (MJ/kg)}
\nomenclature{\(LHV_{F_2}\)}{Transition Fuel LHV (MJ/kg)}
\nomenclature{\(C_{F_1}\)}{Initial Fuel Price (£/tonne)}
\nomenclature{\(\eta_{F_2}\)}{Transition Fuel Boiler Efficiency}
\nomenclature{\(W_j\)}{Thermal Energy required from initial fuel of boiler $j$ (MJ/year)}
\nomenclature{\(J\)}{Set of processes using fuel $F_1$}
\nomenclature{\(G_j\)}{Renewable incentive for transition fuel for boiler $j$ (£/MWh)}

\begin{align}\label{start2}
    \min_{\textbf{G},\mathbf{\phi}} \quad & C_0\left[1-\left(\frac{A_0+\sum_{j=1}^J\phi_j\sum_{j=1}^JW_j/1000LHV_{F_2}}{A_0}\right)^{-L_P}\right] \\
    \text{s.t.} \quad & L_P = {\frac{\log(1-L_r)}{\log(2)}}\\
    & \left[\frac{3.6C_0\eta_{F_2}}{LHV_{F_2}}\right]-G_j - 3.6\frac{C_{F_1}}{LHV_{F_1}} \leq 0 \quad j=1,\dots,J\\
    & \phi_j-\frac{W_i}{\sum_j^J E_i} \leq 0 \quad j = 1,\dots J\\
    & \left[1-\left(\frac{A_0+\sum_{j=1}^J\phi_j\sum_{j=1}^JW_j/1000LHV_{F_2}}{A_0}\right)^{-L_P}\right] - M_U \leq 0 
 \label{end2}
\end{align}

In Eqs. \ref{start2} to \ref{end2}, vectors $\mathbf{G}$ and $\mathbf{\phi}$ represent the fuel grant and market share associated to each process respectively.
The set of processes using fuel is represented by $J$ with individual processes indexed by $j$.
Scalars $LHV_{F_1}$ and $LHV_{F_2}$ are the lower heating value of the initial fuel and alternative fuel respectively. 
$\eta_{F_2}$ is the transition fuel boiler efficiency, and $W_j$ is the thermal energy required from initial fuel of boiler $j$.
Similar to the first case, $M_U$ is a parameter representing the market share limit of alternative fuel and is dictated by existing policy. 
The robust formulation is obtained in a similar manner to the proceeding case study.

\subsection{Cutting-Set Robust Optimisation}
As the class of problems outlined are generally nonlinear, and potentially non-convex due to factors such as learning rates, reformulation of the problem becomes impractical.
Rather than make simplifying assumptions within the model to enable reformulation, we take the view that for this case, maintaining the original model and applying an approximate approach for robust optimisation to immunise solutions against uncertainty provides a greater value. 
To solve the robust market potential assessment problem we apply the Blankenship \& Faulk algorithm \citet{Blankenship1976}. 
Algorithm \ref{algorithm} presents this approach.
\begin{algorithm}[htb!]
\caption{Blankenship \& Faulk}
\label{algorithm}
\KwIn{Objective, $f(\mathbf{x})$, $I$ Semi-infinite constraints $g_0(\mathbf{x},\mathbf{u}),\dots,g_I(\mathbf{x},\mathbf{u})$, Uncertainty-set $\mathcal{U}$, Nominal parameters $\mathbf{u}_0$, Tolerance $\epsilon$}
\KwOut{Robust solution $\mathbf{x}^*$}
\tcp{Initialise cuts for each constraint}
$C_i\leftarrow \{\mathbf{u}_0\} \; i=1,\dots,I$ \leavevmode\newline
\While{True}{
\tcp{Flag problem as currently robust}
    $R = True$ 
\tcp{Solve upper level problem with parameter values from current cutting-set}
$\mathbf{x}^*\leftarrow \mathop{\mathrm{argmin}}_{\mathbf{x}\in\mathcal{X}} \{f(\mathbf{x})\;|g_i(\mathbf{x},\mathbf{u})\leq0, \; \forall \mathbf{u} \in C_i, \; i=1,\dots,I\}$ \leavevmode\newline
\For{i=1,\dots,I}{
\tcp{Maximise constraint violation of current solution}
$\mathbf{u}^*\leftarrow \mathop{\mathrm{argmax}}_{\mathbf{u}\in \mathcal{U}} g_i(\mathbf{x}^*,\mathbf{u})$ \newline
\tcp{If constraint can be violated by a set of parameters...}
\If{$g_i(\mathbf{x}^*,\mathbf{u}^*) > \epsilon$}{
\tcp{Add worst-case parameters to constraints cutting-set}
$C_i \leftarrow C_i \cup \mathbf{u}^*$
\tcp{Flag solution as not robust}
$R = False$}}
\If{$R = \text{True}$}
{\Return $\mathbf{x}^*$}}
\end{algorithm}
This method allows us to maintain the original form of the market potential assessment model, important for considering nonlinear terms under uncertainty such as learning rate.
Algorithm \ref{algorithm} provably converges to the robust solution of a problem, under the assumption that the lower-level problems are solved globally. 
This ensures that robust constraints are not violated, regardless of the value of $\mathbf{u}$ within $\mathcal{U}$.
As global optimisation solvers do not scale well for large problems, we do not solve lower levels globally, and instead, use a local solver with a multi-start heuristic to solve pessimisation sub-problems.
While this provides no guarantees on the optimality of the subproblems, it gives us tractability in solving these real-world formulations.

The subproblems solved between lines 5 and 11 in Algorithm \ref{algorithm} may be solved in parallel enhancing the performance of the algorithm.
In some large problems, the number of constraints added to the upper-level problem becomes unsustainable, and redundant constraints must be `dropped' from the upper-level problem to ensure its tractability. 
However, in the current work, we find this unnecessary. 

\begin{figure}[htb!]
    \centering
    \includegraphics[width=\textwidth]{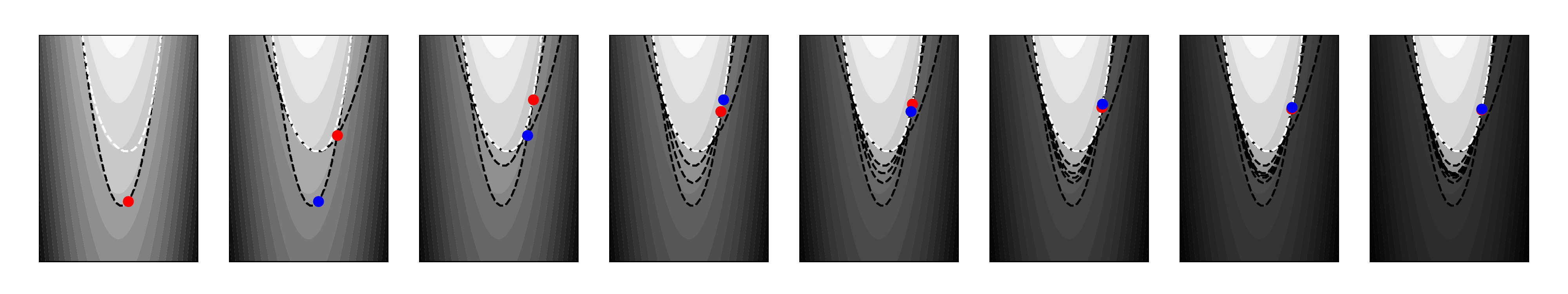}
    \caption{The Blankenship-Faulk algorithm applied to a two dimensional test problem with a single uncertain constraint containing a single uncertain parameter. The red point indicates the current solution to the upper level problem, the blue point indicates the previous solution. An approximation to the robust feasible region is shown in white. The algorithm converges to a tolerance of $1\times10^{-4}$ in 8 iterations.}
    \label{bf_vis}
\end{figure}

For illustration purposes, Figure \ref{bf_vis} demonstrates the behaviour of the Blankenship \& Faulk algorithm applied to a 2 dimensional robust optimisation problem with a single semi-infinite constraint.

The algorithm converges to robust optimal solution in 8 iterations. 
The final semi-infinite constraint is approximated using 8 finite constraints.

\subsection{Reducing Conservatism}

To reduce conservatism in robust optimisation, most commonly, the size or shape of the uncertainty set is changed. 
By applying an ellipsoid shaped uncertainty set, with variable radius, the level of conservatism can be changed and investigated. 
Particularly attractive from a policy-making standpoint is the conversion of ellipsoid-based uncertainty set, to equivalent chance constraint resulting in a probability of policy feasibility. 
This conversion can be made analytically in the case when $g(\mathbf{x},\mathbf{u})$ is linear.
However, as we maintain the nonlinear form of the market potential assessment problem this analytical assumption may be made weaker. 

\subsection{Computational Implementation}

The methodology and case studies are implemented in Python using the Pyomo modelling language \citep{hart2011pyomo} in addition to in-house parallelised cutting-planes code. 
Due to the hardware-specific saving in parallelising the solution to subproblems, we decide to run all case studies on a single node with 64 CPUs. 
This enables the solution of 64 pessimisation subproblems in parallel, greatly enhancing the time to obtain the results. 
The implementation and code to produce all figures can be found at \url{https://github.com/optimaL-pse-lab/rmpa}. 

\section{Case Study and Results}

Case studies of topical alternative technologies and fuels to decarbonise manufacturing are considered in this work. 
Also two typical typologies of policy instruments are designed. 
The first case study is carbon capture, utilisation and storage (CCUS) adoption in the iron and steel industry, with carbon tax as the policy instrument, and the second is hydrogen adoption for energy provision in the chemical industry with incentives as the policy instrument. 

\subsection{Alternative Technology Adoption: CCUS in Iron and Steel Production}

Carbon capture and storage is crucial for emissions abatement in iron and steel production \citep{IEAGHG,Budinis2018}.
CCUS is crucial to achieving carbon neutrality targets, however, the high capital, operation, and maintenance costs as well as the large uncertainties of oil or methanol prices hinder its development.
Integrated steel making using a blast furnace and basic oxygen furnace are the most common routes to make steel \citep{Sun2020}. 
In this case study we investigate the market affects on the uptake of carbon capture and storage within the iron and steel industry.
The policy instrument designated as decision variables is the carbon tax.
The objective is to minimise the difference between the existing cost of capture, and the new cost of capture determined from increased demand in CCUS due to policy intervention.
Equations \ref{con} to \ref{end} describe the problem.
The final nominal problem contains 66 constraints, 64 variables, and 36 uncertain parameters.
The data collected for this research includes the cost of capture for CCS technologies applicable to integrated steel plants, the emissions reduction potential of these technologies, the number of integrated steel plants in the EU, the location of each plant, the steel production capacity of the plants, and the carbon intensity of steel. 
A post combustion capture from blast furnace with 63\% capture rate using 3\% MEA solvent has been demonstrated for use in integrated steel plants with TRL 6 – 8, though this capture rate is generally not precisely known. 
Post-combustion involves removal of carbon dioxide from flue gas generated from a combustion process. 
It is usually carried out through the chemical or physical absorption of carbon dioxide into a solvent.  
In this case study we present 32 iron and steel plants across the EU using MEA Post Combustion capture from blast furnace \citep{Stolaroff2005} with an existing cost of capture ranging from 59.6- 116.3\$/t.

\begin{figure}[htb!]
    \centering
    \includegraphics[width=0.8\textwidth]{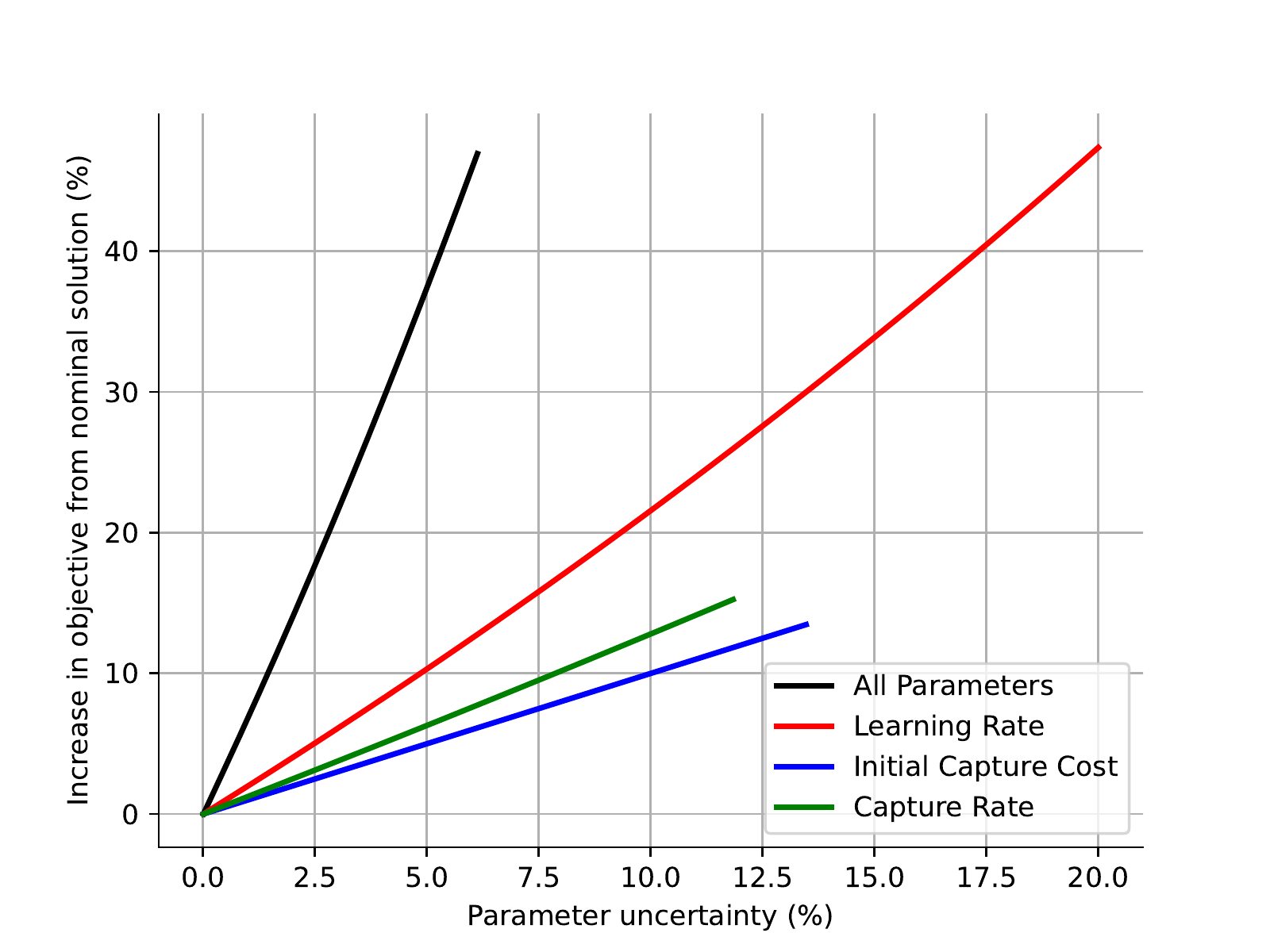}
    \caption{Results of the robust optimisation of the iron and steel case study with a Cartesian product of intervals uncertainty set. The black line demonstrates the increase in objective value when assuming all uncertain parameters can take values of plus or minus the parameter uncertainty level. The red, green and blue lines demonstrate the increase in objective value when just the named parameter is allowed to vary. }
    \label{box}
\end{figure}

\begin{figure}[htb!]
    \centering
    \includegraphics[width=0.8\textwidth]{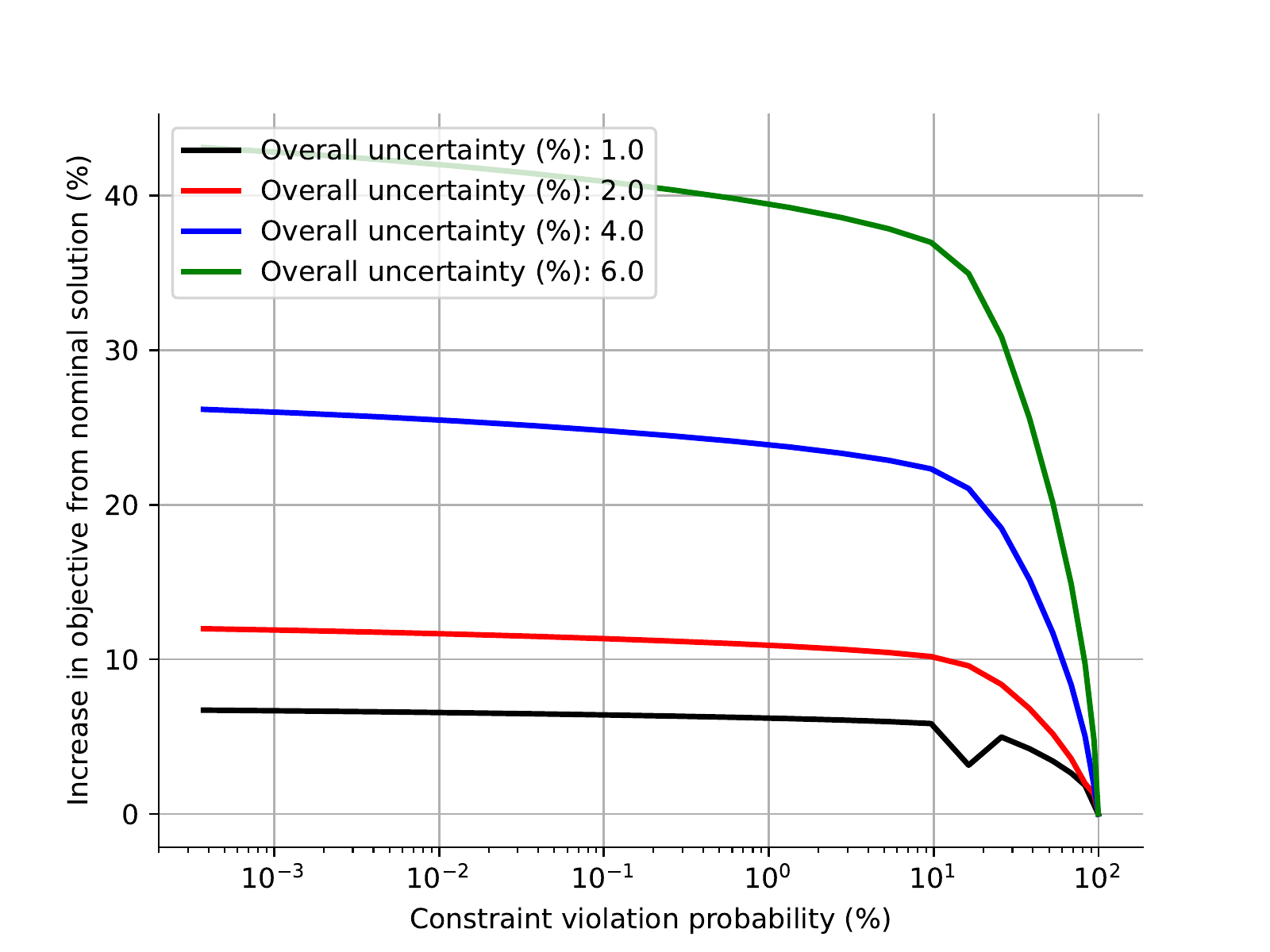}
    \caption{Results of the robust optimisation of the iron and steel case study with an ellipsoidal-based uncertainty set, having converted $\Omega$ to a probability of constraint violation under stated assumptions. }
    \label{e_res}
\end{figure}

To first investigate the effect of uncertainty on the CCUS in iron and steel market potential assessment problem, a Cartesian product of intervals uncertainty set, that defines a hypercube of potential parameters, was applied.
The size of the intervals that specify this uncertainty set were varied across a range of percentage values representing deviations from nominal values. 
Initially these intervals are changed in parallel, representing overall levels of problem uncertainty. 
Subsequently, to identify the effect of each individual parameter's uncertainty on robust policies, individual parameters were considered uncertain and their respective intervals changed.
In this scenario all other parameters take on their nominal values.
Figure \ref{box} demonstrates the percentage increase in optimal objective value (the effectiveness of a robust policy) when considering different levels of parameter uncertainty within a box uncertainty set. 
Figure \ref{box} demonstrates that when all parameters that take on approximately 5\% uncertainty, the robust policy results in an increase in objective value of approximately 40\%.
At 6\% overall uncertainty the problem becomes robustly infeasible, and no set policy interventions can be guaranteed to be feasible solutions to the problem.
This behaviour is expected as this type of uncertainty set is known to be the most conservative.
A hypercube allows for each parameter to take on its worst case value, which rarely happens in practice, particularly in scenarios with a large number of uncertain parameters due to the curse of dimensionality. 
When varying individual parameters the initial capture cost is least sensitive to policy effectiveness. 
However, when the the initial capture cost reaches an interval representing 14\% uncertainty the problem becomes robustly infeasible. 
The learning rate is the most sensitive of the parameters investigated with respect to optimal objective value.
This indicates that focusing on reducing the uncertainty in the learning rate of carbon capture and storage will result in the potential for more effective, robust policies. 
Immunising solutions against a learning rate uncertainty of 20\% which is not untypical provides optimal solutions which are 45\% worse than nominal values. 
These results demonstrate the importance of considering uncertainty within the market potential assessment problem, whilst also highlighting that applying a basic uncertainty set consisting of upper and lower bounds provides extremely conservative solutions that are often infeasible for realistic values of uncertainty.
To provide less conservative solutions an ellipsoidal-based uncertainty set was applied to the robust market allocation problem.
Different formulations are created based on parameter uncertainty level.
The ellipsoid radius $\Omega$ is varied for each formulation. 
Under the assumption that normalised uncertain parameters take values between -1 and 1, have a mean of 0, and are independent of each other: the equivalent percentage chance of constraint violation $\epsilon$ is as follows \citep{BEN09}
\begin{align}
    \epsilon = \exp\{\-\Omega/2\}.\label{cc}
\end{align}
To provide policy-makers more interpretable metrics in order to trade off levels of uncertainty, we vary $\Omega$ to reduce conservatism in robust policies, and subsequently transform $\Omega$ to $\epsilon$ using Eq. \ref{cc}.
Figure \ref{e_res} demonstrates the results of solving the CCUS in iron and steel robust market potential assessment problem, for different uncertainty sets corresponding to probabilities of constraint violation.
Different values of overall uncertainty were investigated.
However, specific bounds of parameters may be used if this information was available.

\begin{figure}[htb!]
    \centering
    \includegraphics[width=0.66\textwidth]{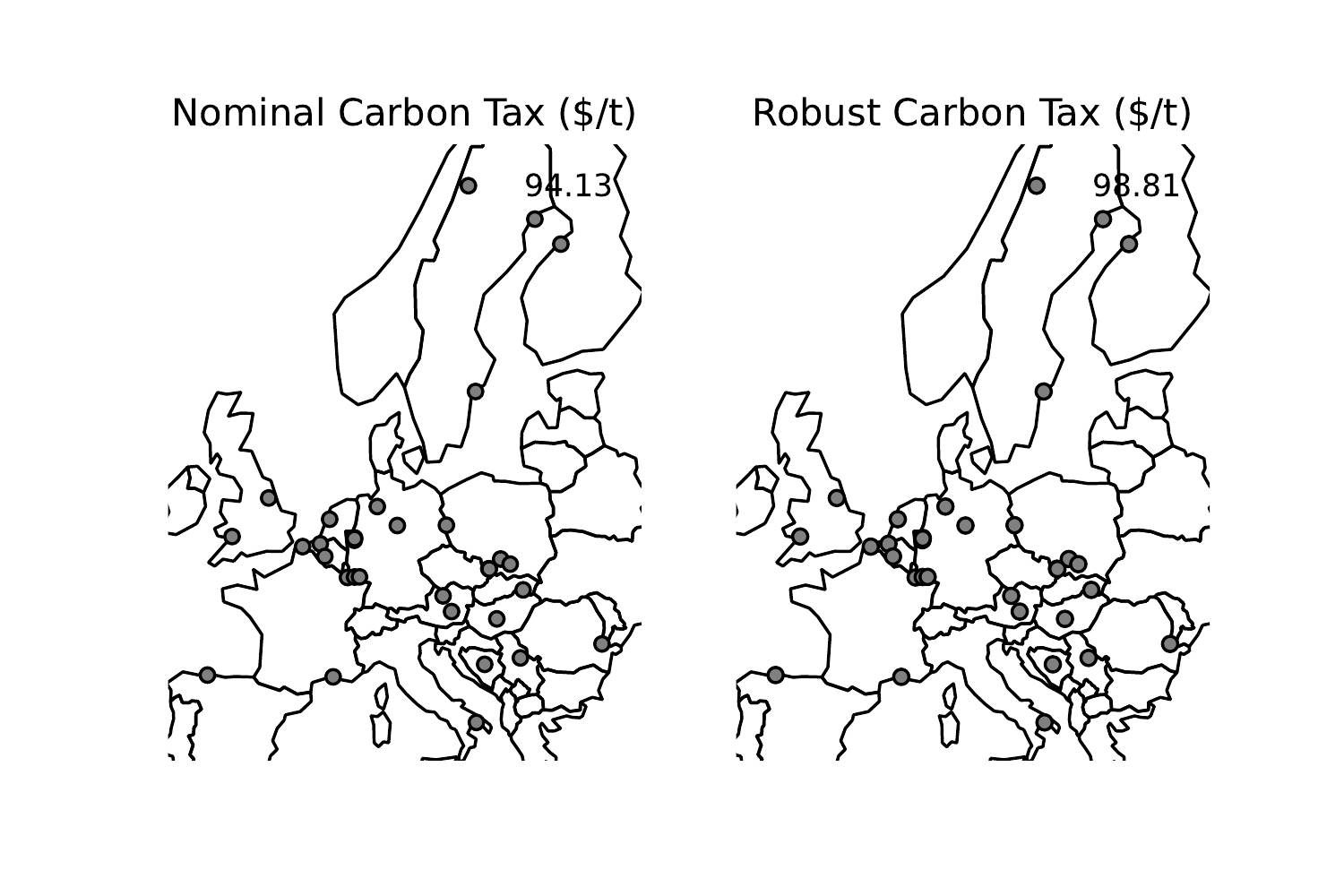}
    \caption{Nominal and robust levels of carbon tax allocated to iron and steel plants across the EU geographic region. The optimisation formulation allocates the same level of carbon tax to all plants, with robust allocation being slightly higher.}
    \label{carbon_geo}
\end{figure}

The smallest ellipse, reducing to a single point of nominal parameters, corresponds to a certain (100\%) constraint violation appears on the right hand side of the graph and is equal to the nominal policy. 
As the constraint violation probability decreases moving left along the x-axis, the optimal objective value increases.
Under the scenario in which parameters take on 4\% of uncertainty from nominal values (blue line) the optimal objective value increases more-or-less linearly with exponentially decreasing constraint violation probability. 
The increase in optimal objective from 100\% chance of violation to 10\% chance is generally large. 
This can be interpreted as the majority of uncertainty being accounted for, with any decreases in violation probability providing less effect on optimal objective value.
In other words, diminishing returns for robust-ifying this policy.
We argue that to properly communicate uncertainty, and robust policy solutions, the graph shown in Fig. \ref{e_res} should be made directly available to policy-makers, in order to support decision making for market potential assessment and policy problems. 
In order to gain a sense of what specific policy differences occur after accounting for robustness, a single scenario was considered. 
All parameters were assumed to have levels of uncertainty corresponding to 5\% of nominal values. 
An ellipsoidal uncertainty set was applied with radius $\Omega=3.7$ corresponding to approximately 0.1\% chance of constraint violation, a value deemed representative of that from a typical decision maker seeking to immunise against uncertainty. 
Figure \ref{carbon_geo} demonstrates the levels of carbon tax allocated to each iron and steel plant across Europe for the nominal and robust problem. 

\begin{figure}[htb!]
    \centering
    \includegraphics[width=\textwidth]{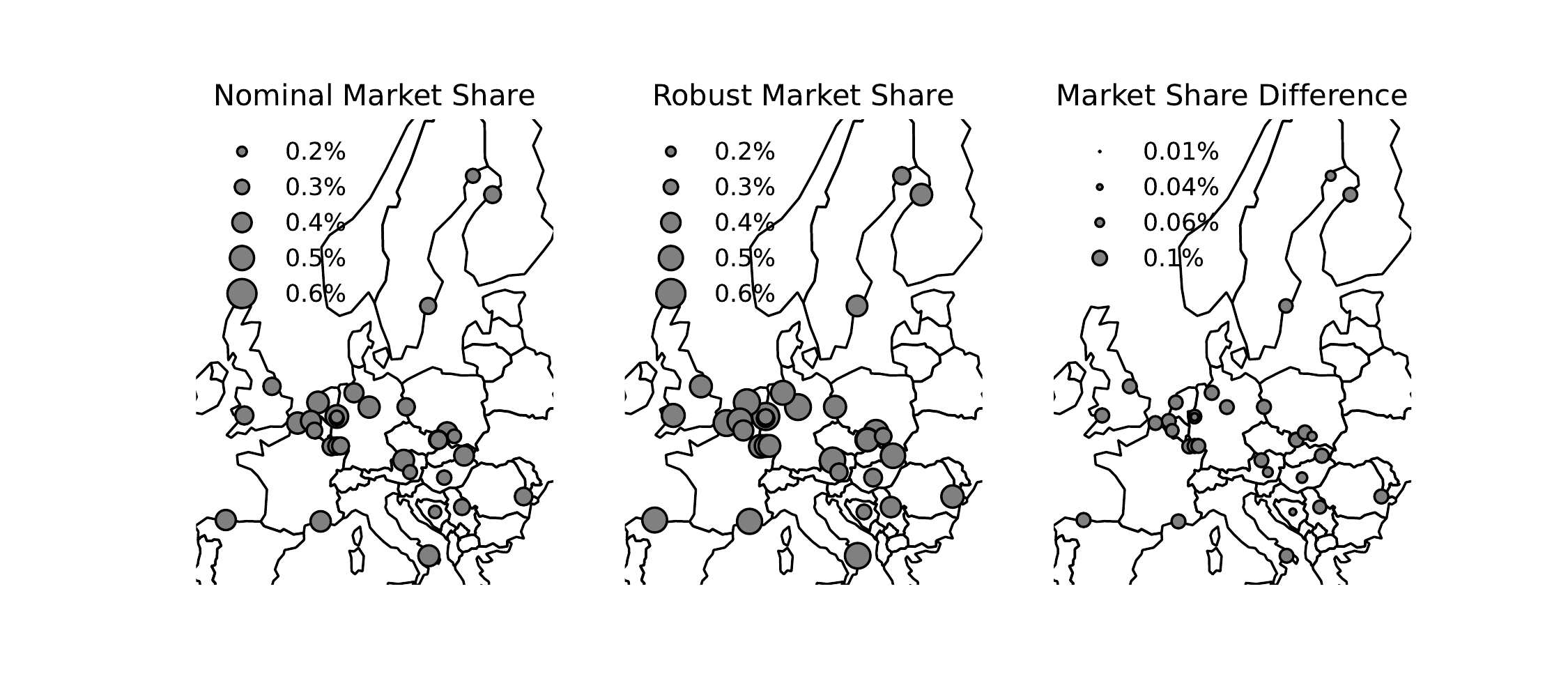}
    \caption{Nominal and robust levels market shares allocated to iron and steel plants across the EU geographic region.}
    \label{market_geo}
\end{figure}

The level of carbon tax allocated was the same for all plants in both problem formulations, with plants being allocated a slightly higher carbon tax in the robust scenario. 
Figure \ref{market_geo} demonstrates the market share allocated to each iron and steel plant across Europe for the nominal and robust problem. 

\begin{figure}[htb!]
    \centering
    \includegraphics[width=0.8\textwidth]{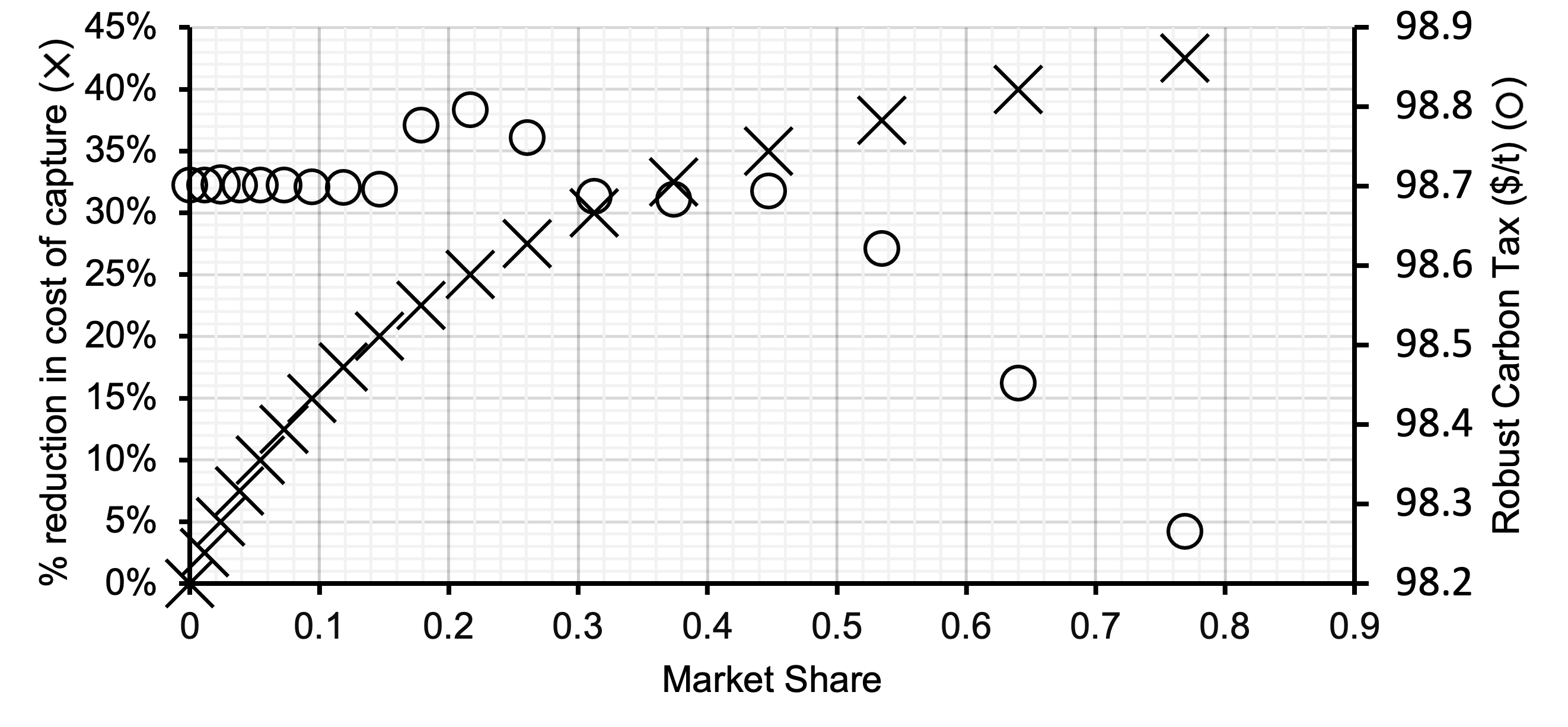}
    \caption{The percentage reduction in the cost of carbon capture and level of carbon tax required to sustain different overall market shares of CCS. As market share increases, carbon tax required to sustain a feasible policy decreases due to learning effects. Likewise, the cost of capture reduces.}
    \label{ct_ms}
\end{figure}

Specific differences in nominal and robust policies can be observed, where a larger market share of CCS is allocated across the fleet of iron and steel plants.
Figure \ref{ct_ms} demonstrates the hypothesis behind the robust market potential assessment for technology adoption in the manufacturing sector. 
The minimum required market share of CCS $M_U$ is varied from decimal valued 0 to 1. 
This shows that the cost of capture can reduce with increase demand due to policy intervention from the carbon tax. 

As the market share required increases from 0 to 1, the percentage reduction in the cost of carbon capture increases.
When approximately 75\% of CO\textsubscript{2} emissions are accounted for with CCS, the cost of carbon capture has the capacity to be reduced by around 43\%. 
Overall levels of carbon tax required to stimulate this demand is less, as market effects increase the demand as opposed to policy incentives. 

\begin{figure}[htb!]
    \centering
    \begin{subfigure}[b]{0.475\textwidth}
    \centering
    \includegraphics[width=0.95\textwidth]{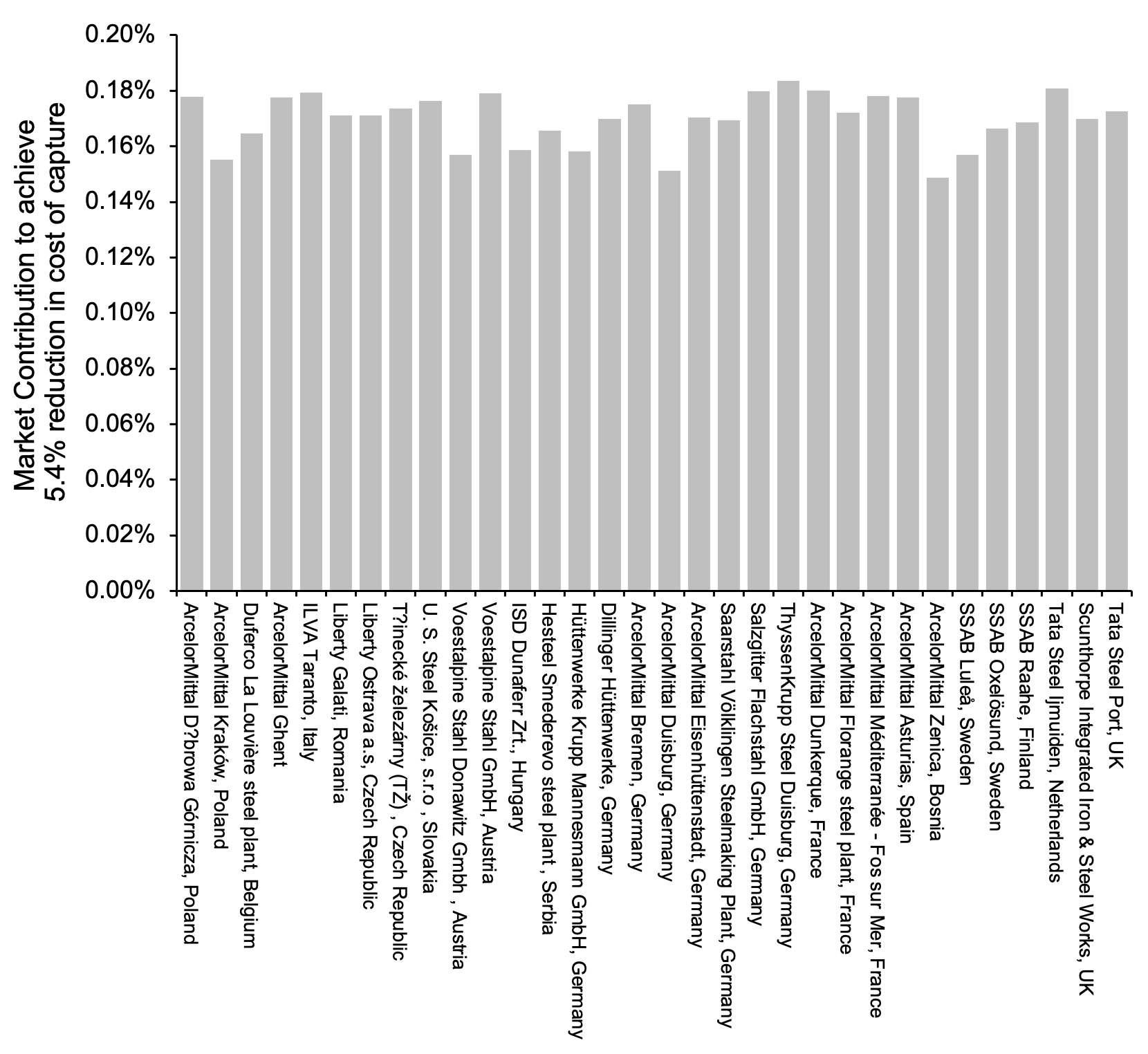}
    \caption{Robust market contribution of each iron and steel plant subject to a minimum 5.4\% reduction in the cost of capture.}
    \end{subfigure}
    \begin{subfigure}[b]{0.475\textwidth}
    \centering
    \includegraphics[width=0.95\textwidth]{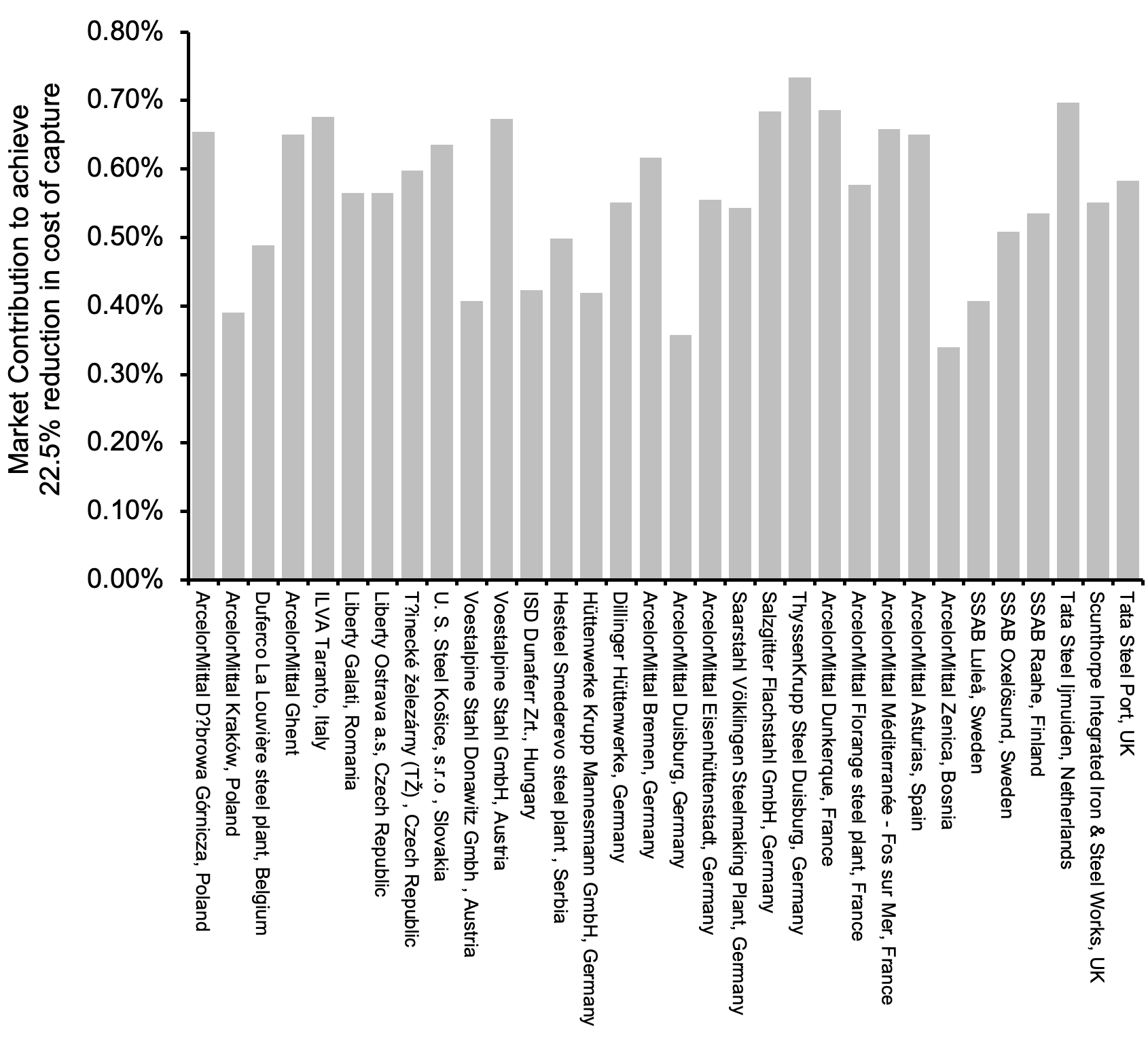}
    \caption{Robust market contribution of each iron and steel plant subject to a minimum 22.5\% reduction in the cost of capture.}
    \end{subfigure}
    \begin{subfigure}[b]{0.475\textwidth}
    \centering
    \includegraphics[width=0.95\textwidth]{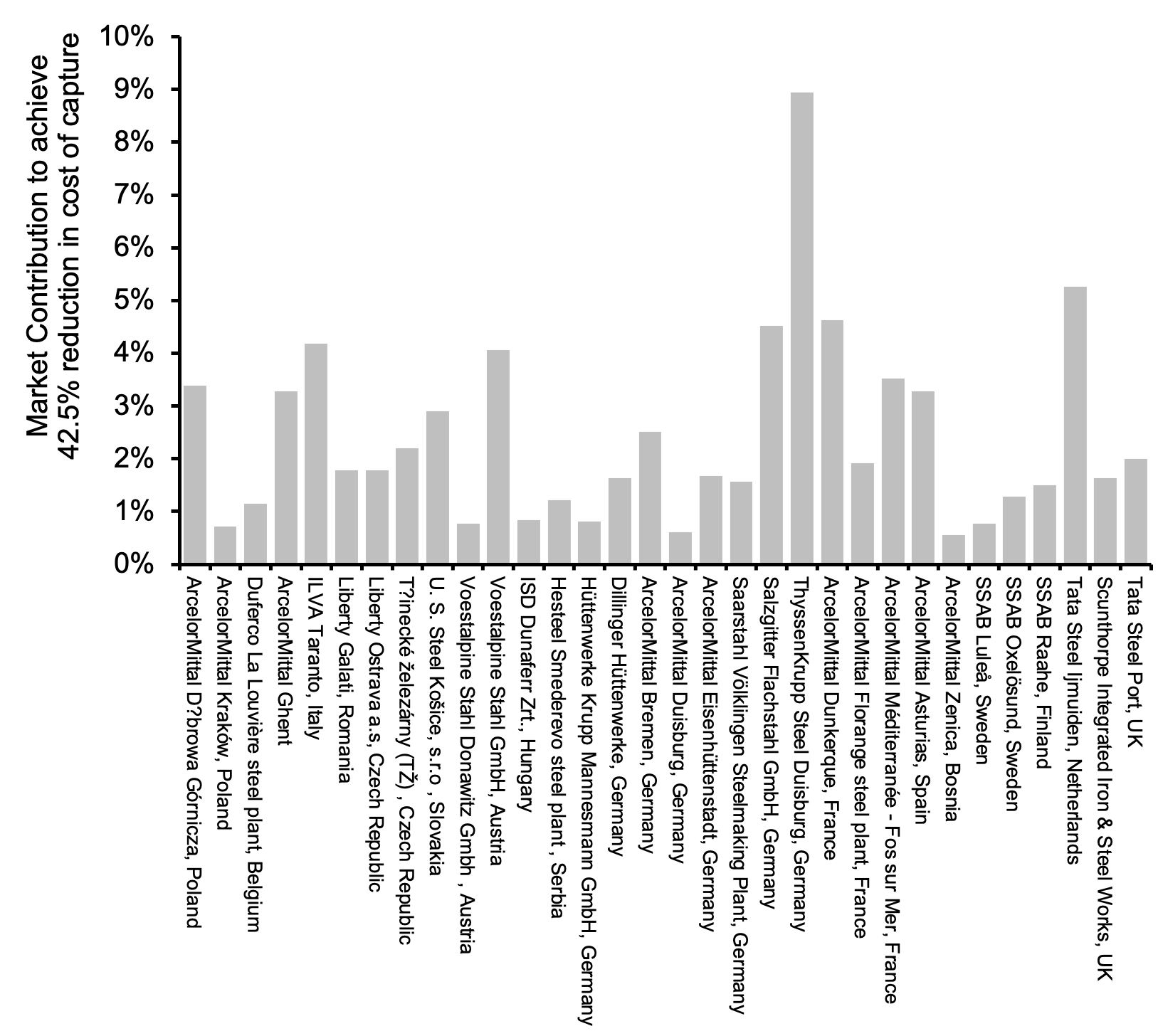}
    \caption{Robust market contribution of each iron and steel plant subject to a minimum 42.5\% reduction in the cost of capture.}
    \end{subfigure}
    \caption{Three separate robust policies, guaranteeing a 5.4\%, 22.5\% and 42.5\% reduction in cost of carbon capture.}
    \label{ms}
\end{figure}

Figure \ref{ms} demonstrates the robust market share of CCUS in each iron and steel plant to achieve 5.4\%, 22.5\% and 42.5\% reduction in cost of capture. 
Demonstrating how the robust market potential assessment formulation accounts for the heterogeneous nature of each plant. 

A greater market penetration is desired, and a more heterogeneous robust policy is dictated. 
This may highlight the need for step-wise policies in future manufacturing policy design in order to minimise inequalities and ensure fairness. 
In addition there is scope for the future investigation of `fair' robust policies for technology adoption. 

\subsection{Alternative Fuel Adoption: Fuel Switching to Blue Hydrogen}

Replacing natural gas burned for heating using industrial boilers and combined heat and power systems is vital to reduce emissions throughout the chemicals sector. The chemical sector is the highest consumer of natural gas in the UK (23\%).
Raising steam by burning hydrogen produced by a low-carbon means is a proposed approach. 
In this case study we investigate the market potential of hydrogen for use within industrial boilers throughout the UK’s chemical industry. 
The policy instrument designated as decision variable is an incentive provided from every unit of heat produced from hydrogen. 
The efficiency of hydrogen boilers is projected to be similar to high-end modern boilers at approximately 90\%. 
This value is not precisely known, should hydrogen boilers become feasible. 
Likewise, the learning rate of hydrogen production is an unknown quantity, currently projected at approximately 7\% \citep{Lambert2019}. 
In this case study we present 354 boilers across chemical plants based in the UK. 
The objective is to minimise the increase in cost of switching to hydrogen within industrial boilers, such that hydrogen becomes cost competitive when taking into account market effects such as learning rate.  
Equations \ref{start2} to \ref{end2} describe the problem. 
Similar to the previous case study, we make the assumption for the purposes of investigation that all parameters are inherently uncertain.  
In this case, these include the reported energy consumption of each boiler, alongside the learning rate of hydrogen, and hydrogen boiler efficiency. 
The final nominal problem contains 710 constraints, 709 variables, and 1073 uncertain parameters.

Relevant data required for this analysis was the price of the natural gas, and blue hydrogen. 
An average of existing 2019 – 2021 data was used to get an estimate of the hydrogen price. 
This led to blue hydrogen being priced at 1800 £/tonne \citep{bloom,hydro_council,iea}.
To ensure consistency, the average 2019 – 2021 natural gas price was found and set as 25 £/MWh or 292 £/tonne \citep{beis}, these define nominal parameter values subject to uncertainty. 
As boilers formed the market scope, the efficiency was determined based on the \citet{iec4h} report. 
The efficiency is shown to not be majorly impacted depending on whether natural gas or hydrogen was used, thus the efficiencies were kept constant.
For boilers this was assumed to be 90\%. 
The initial data collection used to define the market scope of this study was based on the data available in the Hy4Heat report \citep{iec4h}. 
It contained detailed data regarding the number of sector specific sites including the chemicals sector. 
However, the focus was on sites connected to the $<$7 bar gas network and lacked detailed data about the numbers and capacities of boilers in the sectors. 
Therefore, some inferences were made from the data available to establish the first set of boilers as shall be discussed below. 
The \citet{iec4h} dataset was rather limited as it focused solely on $<$7 bar network sites. 
Hence, to expand the market scope and size, data from the CCC’s Sixth Carbon Budget was also used. 
\citet{cccbudg} contained data about boilers in specific chemicals sites and had calculated how much hydrogen (GWh/yr) would be consumed per device per year if fuel switching occurred.

\begin{figure}[htb!]
    \centering
    \includegraphics[width=0.6\textwidth]{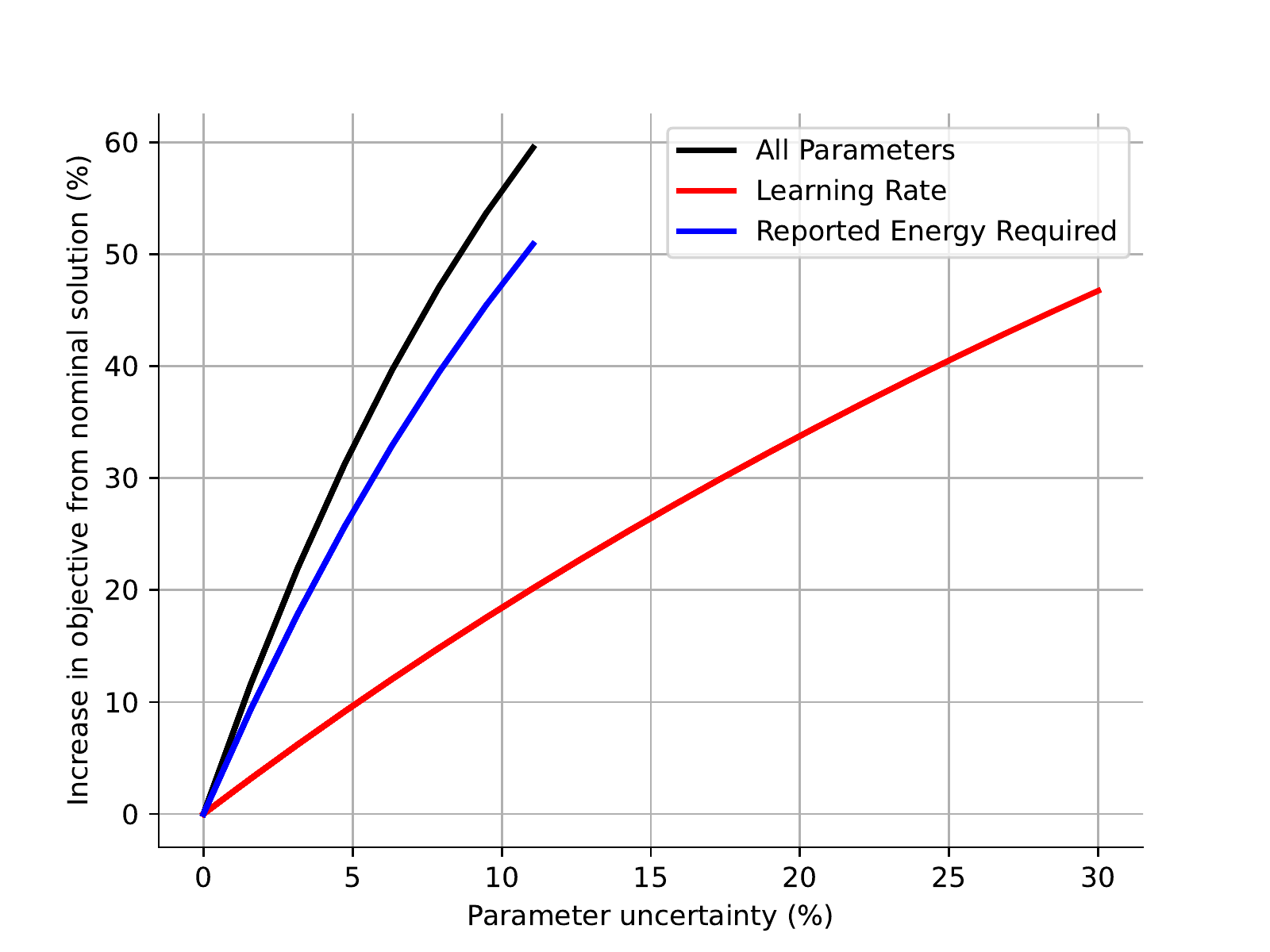}
    \caption{Results of the robust optimisation of the fuel switching study with a Cartesian product of intervals uncertainty set. The black line demonstrates the increase in objective value when assuming all uncertain parameters can take values of plus or minus the parameter uncertainty level. The red and blue lines demonstrate the increase in objective value when learning rate and reported energy required per boiler are deemed uncertain, respectively and independently to all parameters. }
    \label{box2}
\end{figure}

The robust fuel switching problem was solved on a single node consisting of 64 CPU cores and 64 GB of RAM. 
The time taken to solve the nominal problem was 1.1 seconds.
Overall each instance of the robust market allocation problem takes on average 260.1 seconds to locate a robust solution.
The average time to solve the set of 710 pessimisation sub-problems was 128 seconds, accounting for the majority of solution time.
In our experimental set up, without parralelisation of the sub-problems this rises to 223 seconds, roughly doubling the solution time.
Figure \ref{box2} demonstrates the percentage increase in optimal objective value when considering different levels of parameter uncertainty within a box uncertainty set.

The results for the case of box-uncertainty demonstrate that in the fuel switching case study reported energy required per boiler contributes largely to deviations from the nominal solution. 
Considering a 5\% range of reported required energy values across the fleet of considered boilers contributes to approximately 25\% increase in robust policy effectiveness.
This highlights the importance of providing accurate data to policy makers, if effective and robust policies are to be constructed. 
If the quality of provided data is low, then as Fig \ref{box2} demonstrates, any robust solution will effect the operations of each chemical plant to a greater extent than if accurate data were provided.
Overall the fuel switching market allocation problem exhibits more sensitivity to the inclusion of uncertainty than the iron and steel problem. 
Accounting for an uncertain learning rate of hydrogen cost in the problem exhibits similar behaviour to the iron and steel problem.
Optimal robust policies remain feasible for learning rates that have uncertain ranges up to 30\%.
Overall an uncertain learning rate effects optimal robust solutions less than reported energy required. 

\begin{figure}[htb!]
    \centering
    \includegraphics[width=0.6\textwidth]{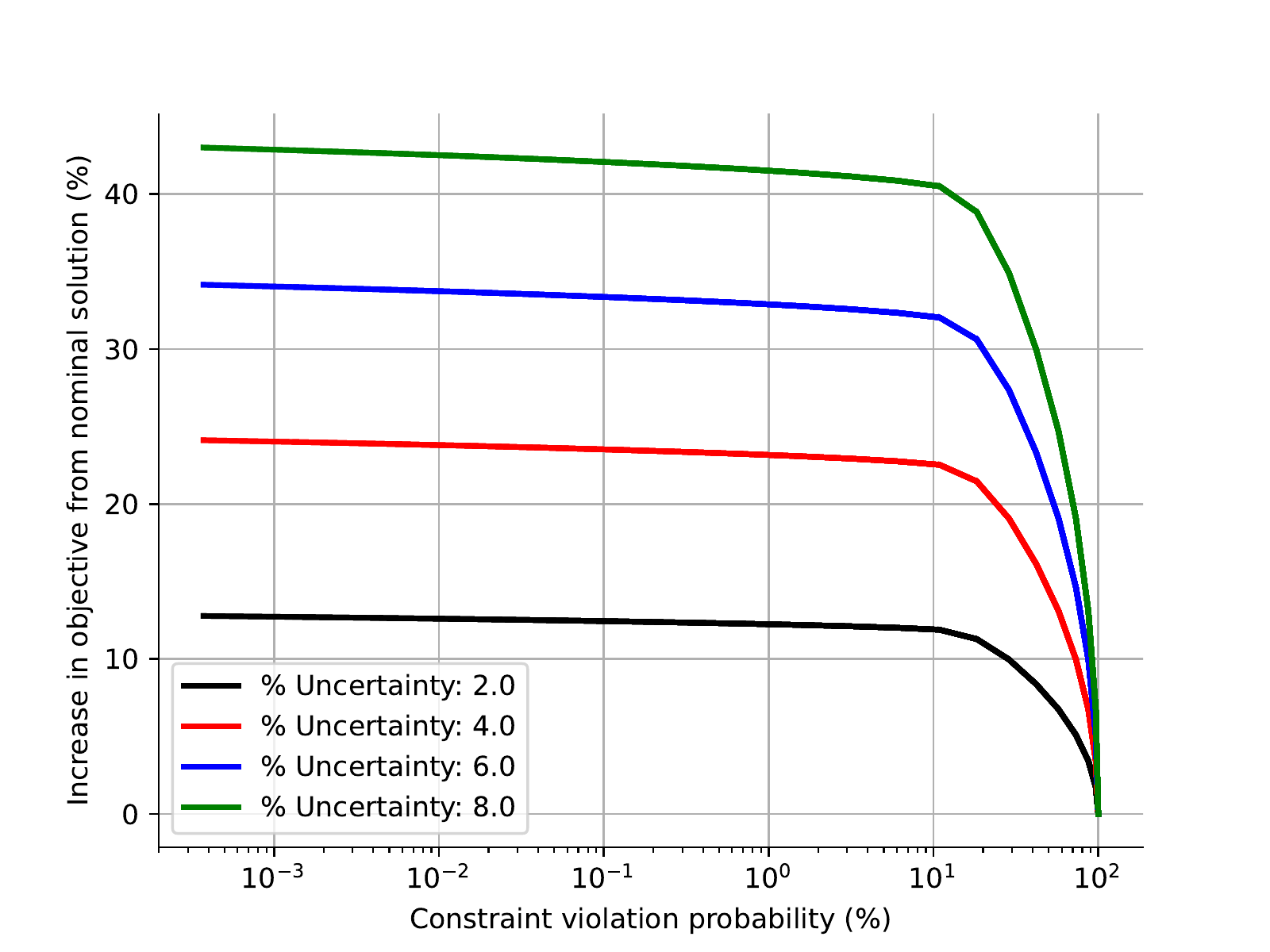}
    \caption{Results of the robust optimisation of the fuel switching case study with an ellipsoidal-based uncertainty set, having converted ellipsoid radius $\Omega$ to a probability of constraint violation under stated assumptions. }
    \label{fuel_switching_prob}
\end{figure}

Under different scenarios of the magnitude of uncertainty of all parameters, the behaviour of robust objective function is similar to the first case study. 
Significant increase in optimal objective function value occurs to immunise a solution against 90\% of uncertainty.
In this case study there is marginal increase from 90\% immunity to 99 or 99.9\% immunity suggesting the potential attractiveness of a fully robust solution. 

\begin{figure}[htb!]
    \centering
    \includegraphics[width=0.6\textwidth]{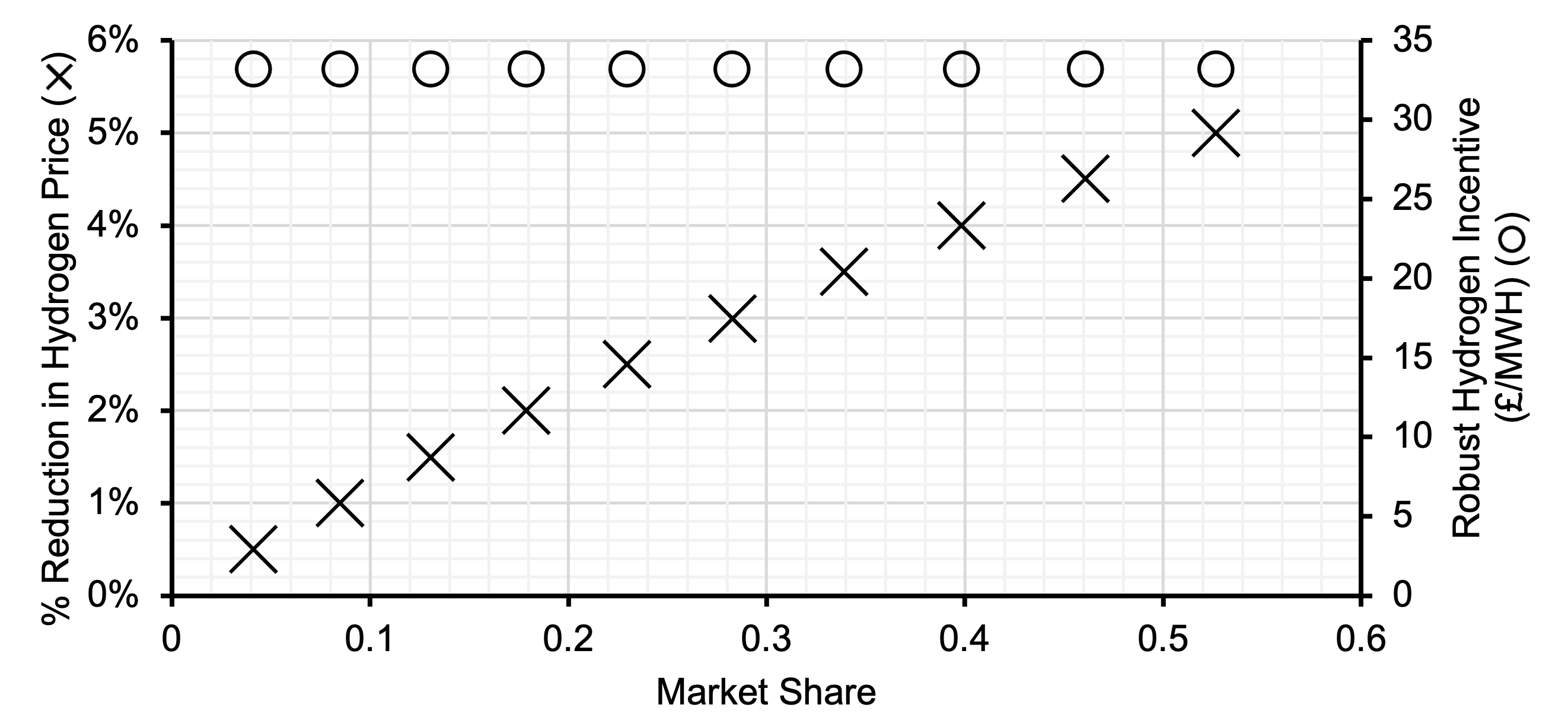}
    \caption{The percentage reduction in the cost of hydrogen and levels of hydrogen incentive required to sustain different overall market shares.}
    \label{fs_ms}
\end{figure}

Figure \ref{fs_ms} demonstrates how the cost of hydrogen can reduce with increase demand due to policy intervention from the hydrogen incentive.

As the minimum market share required of hydrogen is increased, the cost of hydrogen reduces due to technology learning effects. 
Robust hydrogen incentive required to sustain this market share remains approximately constant throughout scenarios. 
This indicates that to guarantee a viable hydrogen economy initial incentives will need to be high regardless of the intended market share. 

\begin{figure}[htb!]
    \centering
    \begin{subfigure}[b]{0.475\textwidth}
    \centering
    \includegraphics[width=\textwidth]{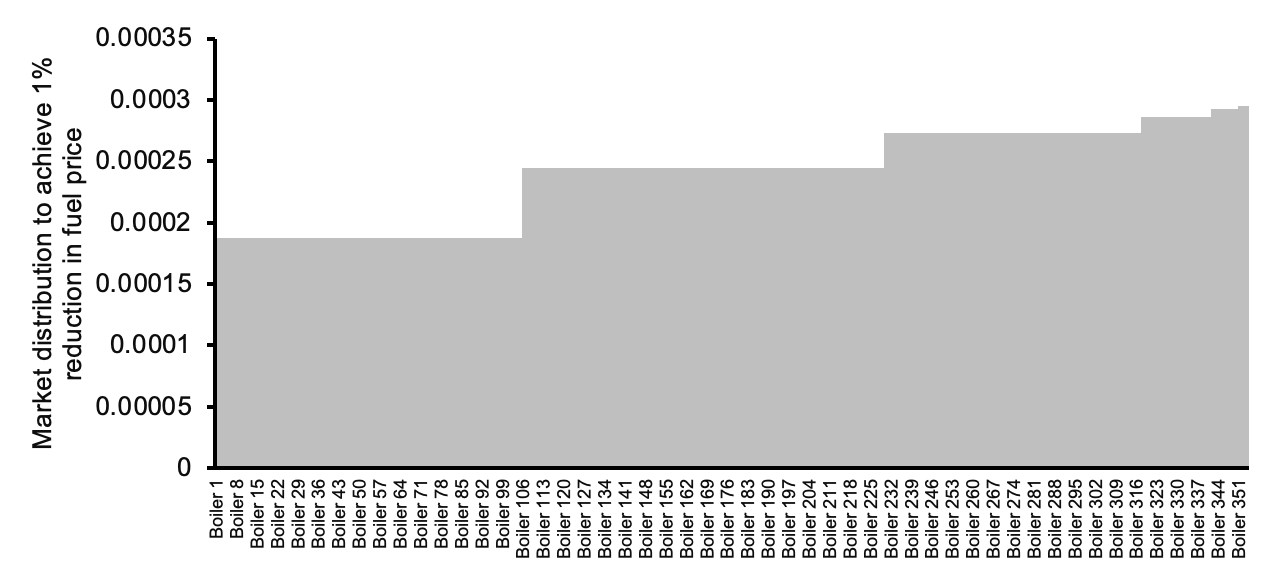}
    \caption{Robust market contribution of each chemical boiler subject to a minimum 1\% reduction in the cost of hydrogen.}
    \end{subfigure}
    \begin{subfigure}[b]{0.475\textwidth}
    \centering
    \includegraphics[width=\textwidth]{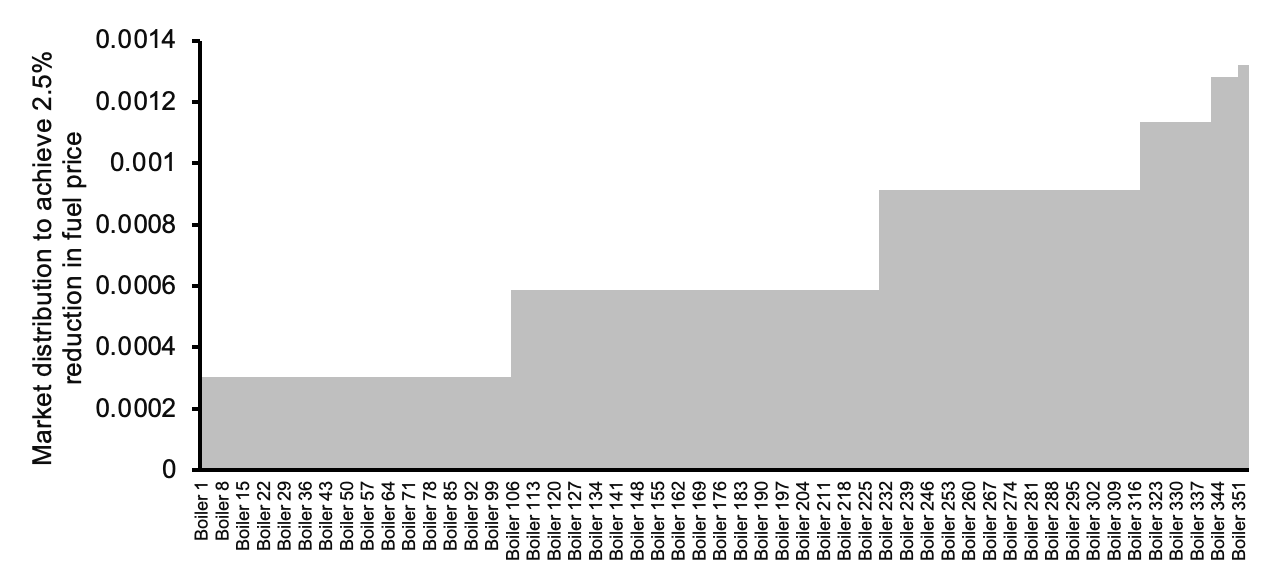}
    \caption{Robust market contribution of each chemical boiler subject to a minimum 2.5\% reduction in the cost of hydrogen.}
    \end{subfigure}
    \begin{subfigure}[b]{0.475\textwidth}
    \centering
    \includegraphics[width=\textwidth]{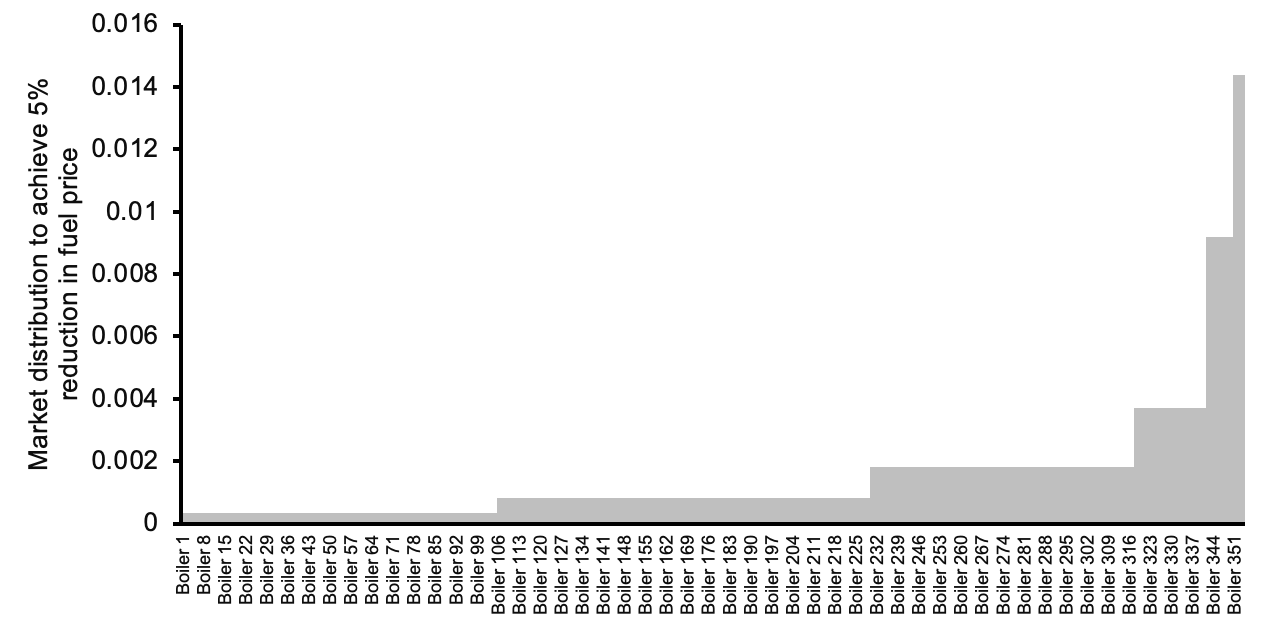}
    \caption{Robust market contribution of each chemical boiler subject to a minimum 5\% reduction in the cost of hydrogen.}
    \end{subfigure}
    \caption{Three separate robust policies, guaranteeing a 1\%, 2.5\% and 5\% reduction in cost of hydrogen.}
    \label{md}
\end{figure}

Figure \ref{md} demonstrates the robust market share of hydrogen across each chemical boiler to achieve 1\%, 2.5\% and 5\% reduction in cost of hydrogen.  

As a larger reduction is required from a robust hydrogen incentive policy, incentives become more targeted. 
Information from Figures such as Fig. \ref{md} can directly enable policy makers to investigate the effect of uncertainty on policy decisions.

\section{Conclusions}

The novel robust market potential assessment model enables for the design of feasible future energy policies to support the uptake of low carbon technologies in the manufacturing sector.  
There is inherent uncertainty associated with aspects of low carbon technologies. 
For example, the learning rates of future low-carbon technologies which dictates how much the cost of a technology will decrease as its market share increases. 
In this paper we formulate the technology adoption problem as a robust optimisation problem, resulting in policies that are immune to uncertain factors. 
The case studies we consider are the potential use of carbon capture and storage for iron and steel production across the EU, and the transition to hydrogen from natural gas in steam boilers across the chemicals industry in the UK. 
Our contributions are three-fold.

\begin{itemize}
\item Development of a robust market potential assessment model for technology adoption in the manufacturing sector. The model is capable of designing policies to generate sufficient demand to trigger and sustain cost reduction until the market takes over. The cost reduction is linked to the technology learning effects
\item Application of a competitive robust optimisation method, with computational enhancements, enabling direct use on existing future energy policy models. 
This approach benefits from parallelisation and ease-of-use, and we release the code-base for future use under the principles of FAIR data and research.
\item We demonstrate the importance of including uncertainty when including market factors such as technology learning rates as well as reported consumption.  
If efficient policies are to be implemented then we highlight the importance of sharing accurate operational data.
\end{itemize}

We present a methodology for reducing the conservatism of robust future energy policy models. 
We demonstrate how our methodology can result in significantly better solutions that are statistically likely to remain feasible. 
Our results demonstrate the possibility of locating and communicating robust policies for the implementation of low-carbon technologies, as well as providing direct insights for policy-makers into the decrease in policy effectiveness that results from increasing robustness.

\section{Authorship Contributions} \textbf{Tom Savage}: Conceptualisation, Methodology, Software, Investigation, Writing - Original Draft. \textbf{Gbemi Oluleye}: Conceptualisation, Methodology, Investigation, Writing - Original Draft, Supervision. \textbf{Antonio Del Rio Chanona}: Supervision, Methodology, Writing - Original Draft

\section{Acknowledgements}
Tom Savage would like to acknowledge the Imperial College President's Scholarship fund as well as Miguel Angel de Carvalho Servia and Damien van de Berg for proofreading the manuscript.




\bibliographystyle{elsarticle-num-names}
\bibliography{references.bib}

\end{document}